\documentclass[10pt,aps,pra,superscriptaddress,floatfix,nofootinbib,superscriptaddress]{revtex4}
\usepackage{graphicx}
\usepackage[english]{babel}
\usepackage{amsmath}
\usepackage{amssymb}
\usepackage{ulem}
\usepackage{subfigure}
\usepackage{bm}
\usepackage{titlesec}
\usepackage{mathrsfs}
\usepackage{MnSymbol}

\newcommand{\out}{\scriptsize\mbox{out}}
\newcommand{\outf}{\scriptsize\mbox{out-fluc}}
\newcommand{\inn}{\scriptsize\mbox{in}}
\newcommand{\innf}{\scriptsize\mbox{in-fluc}}
\newcommand{\sgn}{\mbox{sgn}}
\newcommand{\bx}{{\mathbf x}}
\newcommand{\by}{{\mathbf y}}
\newcommand{\bz}{{\mathbf z}}
\newcommand{\bw}{{\mathbf w}}

\newcommand{\bk}{{\mathbf k}}
\newcommand{\bv}{{\mathbf v}}
\newcommand{\bE}{{\mathbf E}}
\newcommand{\bB}{{\mathbf B}}

\newcommand{\bK}{{\mathbf K}}
\newcommand{\bS}{{\mathbf S}}
\newcommand{\bF}{{\mathbf F}}
\newcommand{\im}{\,{ \rm Im}\, }
\newcommand{\re}{\,{ \rm Re}\, }

\newcommand{\reg}{\scriptsize\mbox{reg}}
\newcommand{\bG}{{\mathbf G}}

\begin{document}

\title{Nonequilibrium quantum fluctuations of a dispersive medium:\\
Spontaneous emission, photon statistics, entropy generation, and stochastic motion}
\author{Mohammad F. Maghrebi\footnote{Present address: Joint Quantum Institute, NIST and University of Maryland, College Park, Maryland 20742, USA.}}
\affiliation{Center for Theoretical Physics, Massachusetts Institute of Technology, Cambridge, MA 02139, USA}
\affiliation{Department of Physics, Massachusetts Institute of Technology, Cambridge, MA 02139, USA}
\author{Robert L. Jaffe}
\affiliation{Center for Theoretical Physics, Massachusetts Institute of Technology, Cambridge, MA 02139, USA}
\author{Mehran Kardar}
\affiliation{Department of Physics, Massachusetts Institute of Technology, Cambridge, MA 02139, USA}

\begin{abstract}
We study the implications of quantum fluctuations of a dispersive medium, under steady rotation, either in or out of thermal equilibrium with its environment.
A rotating object exhibits a quantum instability by dissipating its mechanical motion via spontaneous emission of photons, as well as internal heat generation. Universal relations are derived for the radiated energy and angular momentum as trace formulas involving the object's scattering matrix. We also compute the quantum noise by deriving the full statistics of the radiated photons out of thermal and/or dynamic equilibrium. The (entanglement) entropy generation is quantified, and the total entropy is shown to be always increasing. Furthermore, we derive a Fokker-Planck equation governing the stochastic angular motion resulting from the fluctuating back-reaction frictional torque. As a result, we find a quantum limit on the uncertainty of the object's angular velocity in steady rotation. Finally, we show in some detail that a rotating object drags nearby objects, making them spin parallel to its axis of rotation. A scalar toy model is introduced in the first part to simplify the technicalities and ease the conceptual complexities; a detailed discussion of quantum electrodynamics is presented in the second part.
\end{abstract}

\maketitle

\section*{Introduction}

Fluctuation-induced phenomena have been widely explored in equilibrium where a global temperature exists and the medium (consisting of one or more objects) is static.  Out of equilibrium, quantum and thermal fluctuations can give rise to a rich set of phenomena. A special case of interest is stationary non-equilibrium where there is a temperature gradient, or a medium in steady motion. Energy radiation, friction and dissipation are among the most common themes in this realm.

Here, we explore both thermal and and dynamic nonequilibrium with an emphasis on the latter. Specifically, when neutral objects are set in motion, they interact with quantum fluctuations in the background environment in a time-dependent fashion which may excite photons from the vacuum and lead to \emph{quantum radiation}. The creation of photons by moving mirrors in one dimension was first discussed by Moore~\cite{Moore70}. Accelerated neutral boundaries radiate energy and thus experience a back-reaction force, or \emph{quantum friction}~\cite{Moore70,Fulling76,Ford82,Jaekel92,Calucci92,MaiaNeto93,Law94, Dodonov95,Barton95, Meplan96,Golestanian99,Silveirinha12}.
Recent experiments mimicking such \emph{dynamical Casimir effects}
rely on quantum interference devices for rapidly changing
boundary conditions of a cavity~\cite{Nori11}.

While a substantial literature is devoted to the dynamical Casimir effect in the context of ideal mirrors with perfect boundary conditions~\cite{Lambrecht96,MaiaNeto96,Montazeri08,MaiaNeto98,Dodonov96,Crocce01}, dielectric and dispersive materials have also been studied in several cases~\cite{Barton93,Calogeracos95}.
In general, the latter is more complicated since a quantum system is usually described  by a Hamiltonian, which is lacking for a lossy system. A path integral formulation is also not trivial since the physical system is out of equilibrium,
necessitating the more complicated formalism developed by Schwinger and Keldysh~\cite{Schwinger61,Keldysh64}; several applications of this formalism to quantum friction is investigated in Refs.~\cite{Fosco07,Mkrtchian13}. Interestingly, dispersive objects experience quantum friction even when they move at a constant relative velocity: Two parallel plates moving laterally with respect to each other experience a (non-contact) frictional force~\cite{Pendry97,Volokitin99}. Non-contact friction is usually treated within the framework of the Rytov formalism which is grounded in application of the fluctuation-dissipation theorem to electrodynamics~\cite{Rytov89}. Recently, it was shown that a quantum analog of Chrenkov effect appears when neutral objects are in relative motion beyond a threshold velocity set by the speed of light inside the medium \cite{Maghrebi13-2}; see also Ref.~\cite{Silveirinha13}.

While a constant translational motion requires at least two bodies (otherwise, it is trivial due to Lorentz symmetry), a single spinning object can experience friction~\cite{Manjavacas10}, a phenomenon closely connected to \emph{superradiance} first introduced by Zel'dovich~\cite{Zel'dovich71}. He argued that a rotating object amplifies certain incident waves, and speculated that this would lead to spontaneous emission when quantum mechanics is considered. In the context of general relativity, the Penrose process provides a mechanism similar to superradiance to extract energy from a rotating black hole~\cite{Penrose69}, which also leads to quantum spontaneous emission~\cite{Unruh74}. This radiation, however, is different in nature from Hawking radiation which is due to the existence of event horizons~\cite{Hawking1975}. One can also find similar effects for a superfluid where a rotating object experiences friction even at zero temperature~\cite{Volovik99}.\footnote{
For another proposal related to Casimir-like forces in a slowly moving superfluid, see Ref.~\cite{Roberts05}.}

In this article we expand on a previous short letter~\cite{Maghrebi12}
dealing with quantum fluctuations of a rotating object.
We treat vacuum fluctuations in the presence of a dispersive object under rotation
exactly, except for the assumption of small enough velocities to avoid complications of relativity, thus going beyond the approximate treatments
of previous works in Refs.~\cite{Zel'dovich86,Manjavacas10}.
By incorporating the Green's function techniques into the Rytov formalism~\cite{Rytov89}, we show that a rotating object spontaneously emits photons.

Since we aim to present in detail computations leading to the results briefly
described in Ref.~\cite{Maghrebi12}, this paper is necessarily heavy in technical content. Given that we also derive a number of new results (pertaining to counting statistics, entropy and stochastic motion),
it is important that the mathematical formalism does not obscure their conceptual
simplicity.
As such, in the remainder of the introduction we summarize the important
results, in the order in which they appear in the main text.

{We consider a solid of revolution rotating around its axis of symmetry at a rate $\Omega$. For the sake of definiteness, we assume that the object is at a finite temperature $T$ immersed in a zero-tem
perature vacuum. Generalization to a finite environment temperature is straightforward. The rotating object is characterized by its scattering amplitude $S$ which, according to the symmetries of the problem, is diagonal in frequency $\omega$ and the angular momentum along the rotation axis (in units of $\hbar$), $m$. We use $\alpha$ as a shorthand for all quantum numbers including $\omega$ and $m$, and denote ${\cal I}_\alpha$ as the operator corresponding to the current, or the number of radiated photons, in a partial wave $\alpha$.
\begin{itemize}
  \item A dispersive object under rotation is unstable due to quantum fluctuations. We show that the object spontaneously emits photons at the rate
\begin{equation} \label{Eq: ave number of photons}
\frac{dN_{\alpha}}{d\omega dt} ={\cal N}_{\alpha}= \langle  {{\cal I}_{\alpha}} \rangle=n(\omega-\Omega m,T) \left(1-|S_\alpha|^2\right),
\end{equation}
where $n(\omega,T)=1/(e^{\hbar\omega/kT}-1)$ is the Bose-Einstein distribution function at temperature $T$. Note the shift in frequency due to rotation. This effect persists at zero-temperature,
\begin{equation}
  \lim_{T\to 0}{\cal N}_{\alpha}=\Theta(\Omega m- \omega)\left(|S_\alpha|^2-1\right),
\end{equation}
pointing to its quantum nature.
The scattering matrix is, in fact, super-unitary, $|S_\alpha|>1$, when $\omega <\Omega m$, hence superradiance.

\item The rate of energy/angular momentum radiation, and heat generation are obtained by integrating the current multiplied by the corresponding quantum number, and can be expressed as Trace formulas. The energy radiation, for example, can be written as
    \begin{align}\label{Eq: intro P-M-Q equations}
  {\cal P}&=\int \frac{d\omega }{2\pi} \, \hbar\omega \, \mbox{Tr} \left[n (\omega-\Omega \hat l_z,T) \left(1- {\mathbb S}{\mathbb S}^\dagger\right)\right].
     \end{align}
   The angular momentum radiation and heat generation can be computed by replacing $\hbar \omega \to \hbar m$ and $\hbar \omega \to \hbar (\Omega m-\omega)$, respectively. The loss of angular momentum manifests itself in a quantum friction torque which opposes the rotation of the object.
   Initially at zero temperature, the object loses energy and angular momentum, and heats up at the same time. The energy conservation is respected as the mechanical energy due to rotation is converted into radiation and heat; see Sec.~\ref{Sec: Radiation} for a detailed discussion.

\item We go beyond the averaged value of the radiation, and compute the fluctuations of the (fluctuation-induced) radiation, i.e. higher moments of the current-current correlators. We find the cumulants of factorial moments as
    \begin{equation}
      \kappa_p=\langle {\hat {\cal I}_\alpha}^p\rangle_{c} = (p-1)! {{\cal N}_\alpha}^p,
    \end{equation}
    with the subscript $c$ indicating the \emph{connected }component of the $p$-point function. Remarkably, the average current, $\cal N_\alpha$, also determines all higher moments of fluctuations.

\item Photon statistics can be derived from the knowledge of higher-moment fluctuations. The probability that $n$ photons are radiated in a mode $\alpha$ is given by
   \begin{equation}
     P_\alpha(n)= \frac{{{\cal N}_\alpha}^n}{({\cal N}_\alpha+1)^{n+1}}.
   \end{equation}

\item As the result of radiation, entropy is increased in the environment.
 The entropy generation can be obtained from photon statistics as
\begin{equation}
  {{\cal S}}\equiv \frac{dS}{dt}=k_B\sumint_\alpha \,\,\left(({\cal N}_\alpha+1)\log ({\cal N}_\alpha+1)-{\cal N}_\alpha\log{\cal N}_\alpha\,\right) .
\end{equation}
The symbol $\sumint$ indicates an integral over frequency as well as a sum over other quantum numbers.
This entropy can be interpreted as the entanglement entropy between the object and the environment consisting of radiated photons; see Sec.~\ref{Sec: Entropy}.

\item A freely rotating object slows down as the result of the quantum friction torque $M$, and also undergoes a stochastic motion due to the fluctuational variance of the torque, $\mbox{Var} M$,
    \begin{align}
      M&=\hbar \sumint_\alpha \, m \, {\cal N_\alpha},\nonumber \\
      {\mbox {Var}}M &= \hbar^2 \sumint_\alpha \, m^2 {\cal N_\alpha}\left({\cal N_\alpha}+1\right).
    \end{align}
  The equation of rotation is then a Langevin equation ($I$ being the moment of inertia),
\begin{equation}
  I \dot\Omega(t) = -   M(\Omega(t)) +\eta(t;\Omega(t)).
\end{equation}
The noise $\eta(t;\Omega(t))$ has zero mean, is independent at different times, and correlated fat equal times via
\begin{align}
  \langle \eta(t;\Omega(t))\rangle&=0, \nonumber \\
  \langle \eta(t;\Omega(t))\, \eta(t';\Omega(t'))\rangle &={ \mbox{Var} M}(\Omega(t))\, \delta(t-t').
\end{align}

Equivalently, a Fokker-Planck equation describes the probability distribution as a function of angular velocity; for a detailed discussion, see Sec.~\ref{Sec. Diffusion}.
Even at zero temperature, we find a quantum limit on how sharply the angular momentum can be defined for a single object in steady rotation,
\begin{equation}
  I \Delta\Omega=\sqrt{{I}\frac{ \mbox{Var}M(\Omega_0)}{{\partial M}/{\partial \Omega_0 }}}\propto \sqrt{\hbar \, I\Omega_0}\,.
\end{equation}
Thus the uncertainty in the angular momentum is proportional to the geometrical mean of $\hbar$ and the object's angular momentum (and not $\hbar$ itself).

\item A rotating object makes nearby bodies orbit around its center and also spin parallel to its rotations axis. For the details, see Sec.~\ref{Sec: test object scalar}.

\end{itemize}

}

Our starting point is the Rytov formalism~\cite{Rytov89} which relates fluctuations of the electromagnetic (EM) field to fluctuating \emph{sources} within the material bodies, and in turn to the material's dispersive properties, via the fluctuation-dissipation theorem. While our main focus is moving objects, we find it useful to first consider thermal non-equilibrium where the object and the environment are at different temperatures.
For the sake of simplicity and clarity, we start with a toy model based on a scalar field in Sec.~\ref{Sec: Scalar field}, and postpone the full discussion of electrodynamics to Sec.~\ref{Sec: EM}. For the reader's convenience, the analogy between scalar and electromagnetic fields, and static vs. rotating medium are summarized in Tables \ref{table:1} and \ref{table:2}.

\section{A toy model: A \emph{dielectric} object interacting with a scalar field}\label{Sec: Scalar field}
We consider a scalar field which interacts with an object characterized by a response, or \emph{dielectric}, function $\epsilon$. The reponse function is in principle a function of both frequency and position, and fully characterizes  the object's dispersive properties. The field equation for this model in frequency domain reads
  \begin{equation}\label{scalar field equation}
  \left(\nabla^2+\frac{\omega^2}{c^2}  \epsilon(\omega,\bx) \right)\Phi(\omega,\bx)=0,
  \end{equation}
  with $\epsilon$ being 1 in the vacuum, and a frequency-dependent function inside the object.

  In order to describe quantum\footnote{We shall refer to the quanta of the scalar field as ``photons.''} (and thermal) fluctuations, one can consider the field as a stochastic entity whose fluctuations are governed by a random \emph{source}. From this perspective, quantum fluctuations are cast into a Langevin-like equation (similar to the random force in the theory Brownian motion). For the electromagnetic field, the Rytov formalism provides such a stochastic formulation~\cite{Rytov89}. We introduce a similar approach for the scalar field theory, the central subject of this section. The field equation coupled to a (random) source $\varrho$ is given by
  \begin{equation}\label{Eq: inhom field eqn static}
  -\left(\triangle + \frac{\omega^2}{c^2} \, \epsilon(\omega, \bx)\right)\Phi(\omega, \bx)=-\frac{i \omega}{c} \, \varrho_\omega (\bx)\,,
  \end{equation}
  where the source satisfies a $\delta$-function correlation function in space
  \begin{equation}\label{Eq: source fluc static}
  \langle \varrho_\omega(\bx) \varrho^*_\omega(\by) \rangle =a(\omega)\im \epsilon(\omega, \bx) \, \delta(\bx-\by),
  \end{equation}
  with
  \begin{equation}
  a(\omega)= 2\hbar \left(n(\omega,T)+\frac{1}{2}\right)= \hbar \, \coth\left(\frac{\hbar \omega}{2k_B T}\right)\,.
  \end{equation}
  Note that source fluctuations are related to the imaginary part of the response function in harmony with the fluctuation-dissipation theorem (FDT).
  At a finite temperature $T$, the Bose-Einstein distribution function $n(\omega,T)=[\exp(\hbar \omega/k_B T)-1]^{-1}$ captures thermal fluctuations; the additional 1/2 is due to quantum zero-point fluctuations.

  The field is related to the source via the Green's function, $G$, defined as
  \begin{equation}\label{Eq: Green's function1}
    -\left(\triangle + \frac{\omega^2}{c^2} \epsilon(\omega, \bx)\right)G(\omega, \bx,\bz)=\delta(\bx-\bz).
  \end{equation}
  In equilibrium (uniform temperature with static objects), the field correlation function is obtained as
  \begin{align}\label{Eq: eq FDT1}
    \langle \Phi(\omega, \bx)\Phi^*(\omega, \by)\rangle
    &= \frac{\omega^2}{c^2}\int \int_{\mbox{\scriptsize All space}}
  d\bz \, d\bw \,G(\omega, \bx,\bz) G^*(\omega, \by,\bw) \, \langle \varrho_\omega(\bz) \varrho^*_\omega(\bw) \rangle \nonumber\\
    &= \frac{\omega^2}{c^2} a(\omega) \int_{\mbox{\scriptsize All space}} d \bz \, G(\omega,\bx,\bz) \im \epsilon (\omega,\bz) \, G^*(\omega,\by,\bz) \nonumber \\
    &= a(\omega) \im G(\omega, \bx,\by).
  \end{align}
  Note that the second line in Eq.~(\ref{Eq: eq FDT1}) follows from $\frac{\omega^2}{c^2}\im\epsilon =-{\im G^{-1}}$ according to Eq.~(\ref{Eq: Green's function1}).
  This equation manifests the FDT by relating \emph{field }fluctuations to the imaginary part of the Green's function. However, Eq.~(\ref{Eq: eq FDT1}) requires the system to be in equilibrium while Eq.~(\ref{Eq: source fluc static}) is formulated locally and makes no assumption about global properties of the system such as overall equilibrium. Therefore, we shall employ Eq.~(\ref{Eq: source fluc static}) to study nonequilibrium systems.

  In the following sections we explore the interplay between geometry, motion, and temperature. While our main interest is the consequences of fluctuations in the context of moving objects, we make a detour to study quantum and thermal fluctuations for a static object. Out of thermal equilibrium, the object is at a temperature different from that of the environment. The techniques we develop in the following section are useful when we consider moving objects in or out of thermal equilibrium. For simplicity, we consider a disk in two-dimensional space; generalization to realistic objects is discussed in the context of electromagnetism.

  In the following, we make the convention that $c=1$ unless stated otherwise.

\subsection{{Field fluctuations for static objects}}\label{Field Fluctuations for Static Objects}
According to the Rytov formalism, field fluctuations are induced by random sources which fluctuate according to the object's local properties (encoded by the imaginary part of the response function) and temperature (through the Bose-Einstein factor). It is then natural to divide the space into the object and the environment (vacuum), and to compute the source fluctuations in each region separately.
\subsubsection{Vacuum fluctuations}
In this subsection we consider field fluctuations due to random sources only in the vacuum. The scalar field is coupled to fluctuating sources outside the object as
\begin{align}
   -(\triangle + \omega^2 \epsilon(\omega, \bx))\Phi(\omega, \bx)=
   \begin{cases}
      0, & |\bx|< R, \\
     -i \omega \varrho_\omega (\bx), & |\bx|> R,
   \end{cases}
\end{align}
with $R$ being the radius of the disk. Source fluctuations, according to the Rytov formalism, are determined by
\begin{equation}
 \langle \varrho_\omega(\bx) \varrho^*_\omega(\by) \rangle =a_{\out}(\omega)\im \epsilon_D(\omega) \, \delta(\bx-\by),
\end{equation}
where the points $\bx$ and $\by$ are outside the object, $a_{\out}$ corresponds to the temperature of the environment, and $\epsilon_D$ represents the response functions in the vacuum. It might seem that this function is 1 and $\im \epsilon_D=0$, hence there are no source fluctuations outside the object. However, even in empty space, we need  sources to give rise to zero-point fluctuations. Indeed, as one has to integrate over infinite volume, the limit of $\im \epsilon_D \to 0$ should be taken with care.
The corresponding field correlation function outside the object is given by
\begin{equation}\label{static out fluc 0}
  \langle \Phi(\omega, \bx)\Phi^*(\omega, \by)\rangle_{\outf}= \omega^2 a_{\out}(\omega) \im \epsilon_D(\omega) \int_{{|\bz|>R}}
  d\bz  \, G(\omega, \bx,\bz)  \, G^*(\omega, \by,\bz).
\end{equation}
Note that the Green's functions are evaluated outside the object. Let $(r,\phi)$ and $(\xi,\psi)$ be the polar coordinates of $\bx$ and $\by$, respectively.
The (retarded) Green's function can be cast as a sum over partial waves in the cylindrical basis as
\begin{equation}\label{Green fn out-out}
  G(\omega, \bx,\by)=  \sum_{m=-\infty}^{\infty} \frac{i}{8}\left(H_m^{(2)}(\omega r)+ S_m(\omega) H_m^{(1)}(\omega r)\right) e^{i m\phi} \,\, H_m^{(1)}(\omega \xi) e^{-im\psi}, \hskip .3in R<r<\xi \,,
\end{equation}
where $H_m^{(1,2)}$ are the Hankel functions of the first and second kind, and $S_m(\omega)$ is the scattering matrix.
Furthermore, we have assumed that the point $\by$ is located at a larger radius from the origin without loss of generality.
In empty space, $S=1$, and we recover the free Green's function as
\begin{equation}\nonumber
  G(\omega, \bx,\by)=  \sum_{m=-\infty}^{\infty} \frac{i}{4}J_m(\omega r) e^{i m\phi} \,\, H_m^{(1)}(\omega \xi) e^{-im\psi}, \hskip .3in r<\xi \,.
\end{equation}
To compute the integral in Eq.~(\ref{static out fluc 0}), one should integrate over $R<|\bz|<\infty$; however, we take the limit that $\im \epsilon_D \to 0$, and only a singular contribution, due to the integral over $|\bz|\to \infty$, survives. We can then safely choose the domain of integration as $|\bz|> r, \xi$. We stress that in the intermediate steps, the argument of the Hankel function should be modified to $\sqrt{\epsilon_D} \, \omega r$ with the limit $\epsilon_D \to 1$ taken in the end. A little algebra yields
\begin{align}\label{static out fluc}
  \langle \Phi(\omega, \bx)\Phi^*(\omega, \by)\rangle_{\outf}= \frac{1}{16}& a_{\out}(\omega) \sum_{m=-\infty}^{\infty} \left(H_m^{(2)}(\omega r)+ S_m(\omega) H_m^{(1)}(\omega r)\right) e^{i m\phi} \times \nonumber \\ &\overline{\left(H_m^{(2)}(\omega \xi)+ S_m(\omega) H_m^{(1)}(\omega \xi)\right) e^{i m\psi}}.
\end{align}
(The bar indicates complex conjugation.) The correlation function is then a bilinear sum over incoming plus scattered waves. In fact, in the absence of the object, this equation reduces to a bilinear sum over Bessel functions
\begin{align}\label{Eq: completeness}
  \langle \Phi(\omega, \bx)\Phi^*(\omega, \by)\rangle_{\scriptsize\mbox{empty space}}&= \frac{\hbar}{2} \left(n(\omega,T)+\frac{1}{2}\right) \sum_{m=-\infty}^{\infty}  J_m(\omega r) J_m(\omega \xi) e^{i m(\phi-\psi)}\nonumber \\
  &=\frac{\hbar}{2} \left(n(\omega,T)+\frac{1}{2}\right) J_0(\omega|\bx-\by|) \nonumber \\
  & = \frac{\hbar}{2} \left(n(\omega,T)+\frac{1}{2}\right) \int_0^{2\pi} \frac{d\alpha }{2\pi}\, e^{i \bk\cdot (\bx-\by)},
\end{align}
where $\bk$ is the wavevector with $|\bk|=\omega$ and $\angle\bk=\alpha$.
Being a complete basis, the Bessel functions can be recast into another basis such as planar waves in Eq.~(\ref{Eq: completeness}). In other words, quantum fluctuations in (empty) space can be written as a uniformly-weighted sum over a complete set of functions. In the presence of the object, vacuum fluctuations are organized into a sum over incoming plus scattered waves as in Eq.~(\ref{static out fluc}).

\subsubsection{Inside fluctuations}
Next we turn to study the source fluctuations inside the object:
\begin{align}
   &-(\triangle + \omega^2 \epsilon(\omega, \bx))\Phi(\omega, \bx)=
   \begin{cases}
     -i \omega \varrho_\omega (\bx), & |\bx|< R, \\
     0, & |\bx|> R,
   \end{cases} \\
    \mbox{with} \hskip .4in &  \langle \varrho_\omega(\bx) \varrho^*_\omega(\by) \rangle =a_{\inn}(\omega)\im \epsilon(\omega,\bx) \, \delta(\bx-\by),
\end{align}
where the sources' arguments are inside the object, and $a_{\inn}(\omega)$ is defined with respect to the object's temperature. Similar to the previous section, the field correlation function for $\bx$ and $\by$ \emph{outside} the object can be computed via Green's functions,
\begin{equation} \label{static in fluc 00}
  \langle \Phi(\omega, \bx)\Phi^*(\omega, \by)\rangle_{\innf}= \omega^2 a_{\inn}(\omega) \int_{{|\bz|<R}}
  d\bz  \, G(\omega, \bx,\bz)  \im \epsilon(\omega,\bz)\, G^*(\omega, \by,\bz),
\end{equation}
where $\epsilon$ is a possibly position-dependent response function. The Green's function in the last equation involves a point
inside and another outside the object. As the two points (inside and outside the object) cannot coincide, the Green's function satisfies a {\it homogeneous} equation, Eq.~(\ref{scalar field equation}), inside and a free (Helmholtz) equation outside. Hence, we can expand the Green's function as
\begin{equation} \label{Green fn in-out 0}
  G(\omega, \bx,\bx')=  \sum_{m=-\infty}^{\infty} \frac{i}{8} \, f_{\omega,m}(r) e^{i m\phi} \,\, \left(A \, H_m^{(1)}(\omega \xi)+B \, H_m^{(2)}(\omega \xi)\right) e^{-im\psi}, \hskip .3in r<R<\xi,
\end{equation}
where the prefactor is chosen for future convenience. Here $f_{\omega,m}(\omega)$ is the regular (at the origin) solution to\footnote{For simplicity, we have assumed that the dielectric function is rotationally symmetric. This assumption is not essential for a static object, but is essential for rotating objects.}
\begin{align}\label{Eq: f inside}
   & -(\triangle + \omega^2 \epsilon(\omega, r))\, f_{\omega,m}(r) e^{i m\phi}= 0, \nonumber \\
   &-(\triangle + \omega^2 ) \, H_m^{(1,2)}(\omega r)e^{i m\phi}=0.
\end{align}
The coefficients $A$ and $B$ and the normalization of the function $f$ are determined by matching the Green's functions approaching a point on the boundary from inside and outside the object
\begin{equation}
  G(\omega, \bx,\by){\mid}_{|\bx|\to R^-}=  G(\omega, \bx,\by){\mid}_{|\bx|\to R^+}.
\end{equation}
Comparing Eqs. (\ref{Green fn out-out}) and (\ref{Green fn in-out 0}), we find
\begin{equation} \label{Green fn in-out}
  G(\omega, \bx,\by)=  \sum_{m=-\infty}^{\infty} \frac{i}{8} \, f_{\omega,m}(r) e^{i m\phi} \, H_m^{(1)}(\omega \xi) e^{-im\psi}, \hskip .3in r<R<\xi,
\end{equation}
where the function $f$ is constrained by continuity equations as
\begin{align}\label{continuity}
  & f_{\omega, m} (R)= H_m^{(2)}(\omega R)+ S_m(\omega) H_m^{(1)}(\omega R), \cr
  & \left[\frac{\partial}{\partial r} f_{\omega, m} (r)= \frac{\partial}{\partial r} \left( H_m^{(2)}(\omega r)+ S_m(\omega) H_m^{(1)}(\omega r)\right)\right]_{r=R}.
\end{align}
In short, the differential equations in Eq.~(\ref{Eq: f inside}) plus the boundary conditions in the last equations and the regularity of $f$ at the origin determine both the function $f$ and the elements of the $S$-matrix. We then expand the Green's function in Eq.~(\ref{static in fluc 00}) in terms of partial waves from Eq.~(\ref{Green fn in-out}). Keeping in mind that $\rho\equiv|\bz|<r,\xi$, we find
\begin{align} \label{static in fluc 0}
  \langle \Phi(\omega, \bx)\Phi^*(\omega, \by)\rangle_{\innf}&= \frac{1}{64}\omega^2 a_{\inn}(\omega) \sum_{m=-\infty}^{\infty}
  H_m^{(1)}(\omega r) e^{-im\phi} \overline{H_m^{(1)}(\omega \xi) e^{-im\psi}} \times \nonumber \\
  &2\pi \int_{0}^{R} d\rho\, \rho f_{\omega, m}(\rho)\im \epsilon(\omega,\rho)\, \overline{f_{\omega, m}(\rho)}.
\end{align}
By virtue of the field equation, the integral in the last line of this equation can be converted to an expression on the boundary of the object: The conjugate of the function $f$ satisfies the conjugated wave equation with $\epsilon \to \epsilon^*$. By subtracting off the conjugated from the original equation, one can see that the integrand is equal to a total derivative. The integral then becomes
\begin{equation}
  \frac{1}{-2i \omega^2} \, W\left(f_{\omega,m}(R),\overline{f_{\omega,m}(R)}\right),
\end{equation}
with $W$ being the Wronskian with respect to the radius. The continuity relations of Eq.~(\ref{continuity}) can be exploited to compute the Wronskian
\begin{equation}
  W\left(f_{\omega,m}(R),\overline{f_{\omega,m}(R)}\right)= -\frac{4i}{\pi R} \left(1-|S_m(\omega)|^2)\right),
\end{equation}
where we used the identity $W\left(H_m^{(1)}(x),H_m^{(2)}(x)\right)=-4i/\pi x$.
Rather remarkably, this equation shows that all the relevant details of the inside solutions $f$ can be encoded in the scattering matrix, i.e. fluctuations inside the object affect the correlation function only through the scattering matrix, $S$. Combining the previous steps, we arrive at the (outside) correlation function due to the inside source fluctuations,
\begin{equation}\label{static in fluc}
  \langle \Phi(\omega, \bx)\Phi^*(\omega, \by)\rangle_{\innf}= \frac{1}{16} a_{\inn}(\omega) \sum_{m=-\infty}^{\infty} \left(1-|S_m(\omega)|^2\right)   H_m^{(1)}(\omega r) e^{im\phi} \overline{H_m^{(1)}(\omega \xi) e^{im\psi}}.
\end{equation}
The correlation function is a bilinear sum over outgoing (first kind of Hankel) functions; this is reasonable as the sources in the object must produce outgoing waves in the vacuum. The coefficient is, however, more interesting: It depends on the scattering matrix through $1-|S|^2$, and vanishes for a non-lossy object, i.e. when the scattering matrix is unitary, $|S|=1$. We shall revisit this point later when we study radiation out of thermal or dynamic equilibrium.

\subsubsection{Thermal radiation}
In this section, we employ the results from the previous sections to compute the radiation out of thermal equilibrium when the object is at rest, though at a temperature $T$ different from that of the environment, $T_0$. But we first show that the equilibrium behavior is consistent with the FDT. At $T=T_0$, the distribution functions $a_{\inn}(\omega)=a_{\out}(\omega)\equiv a(\omega)$ are equal. A sum over Eqs. (\ref{static out fluc}) and (\ref{static in fluc}) yields (for $\bx$ and $\by$ outside the body)
\begin{align}\label{Eq: FDT}
  \langle \Phi(\omega, \bx)\Phi^*(\omega, \by)\rangle
  &= \langle \Phi(\omega, \bx)\Phi^*(\omega, \by)\rangle_{\outf}+ \langle \Phi(\omega, \bx)\Phi^*(\omega, \by)\rangle_{\innf} \nonumber \\
  &= a(\omega) \, \im \sum_{m=-\infty}^{\infty} \frac{i}{8}\left(H_m^{(2)}(\omega r)+ S_m(\omega) H_m^{(1)}(\omega r)\right) e^{i m\phi} \,\, H_m^{(1)}(\omega \xi) e^{-im\psi} \nonumber\\
  &= a(\omega) \, \im \, G(\omega, \bx, \by),
\end{align}
in agreement with the FDT.

Out of thermal equilibrium, the ``Poynting'' vector quantifies the radiation flux from the object into the environment. In our model for the scalar field, the radial component of the Poynting vector is given by
\begin{align}\label{Eq: Poyn vector}
\langle \partial_t \Phi (t,\bx) \partial_r \Phi (t,\bx)\rangle
  = \frac{1}{\pi}\int_{0}^{\infty}d\omega \, \omega \im \langle \Phi(\omega,\bx) \partial_r\Phi^*(\omega,\bx) \rangle.
\end{align}
The total radiation rate is obtained by integrating over a closed surface enclosing the object. We compute the contribution due to inside and outside source fluctuations separately by inserting the corresponding correlation functions in the last equation. The radiated energy per unit time is then
\begin{align}
  {\cal P}_{\innf/\outf}=\pm\frac{1}{4\pi} \sum_{m=-\infty}^{\infty} \int_{0}^{\infty} d\omega \, \omega a_{\inn/\out}(\omega) \left(1- |S_m(\omega)|^2\right),
\end{align}
with the upper (lower) sign corresponding to inside (outside) fluctuations, where we have used the expression for the Wronskian of Hankel functions. Note that the signs indicate that the flux due to the inside sources is outgoing while the vacuum fluctuations induce an incoming flux. In the absence of loss, i.e. when $|S|=1$, there is no flux in either direction since the object lacks an exchange mechanism with the environment. In equilibrium, detailed balance prevails and there is no net radiation. One can also see that the reality of the correlation function in Eq.~(\ref{Eq: FDT}) guarantees that the corresponding Poynting vector in Eq.~(\ref{Eq: Poyn vector}) vanishes. Out of thermal equilibrium, the total radiation to the environment is given by
\begin{equation}\label{Eq: Kirchhoff}
  {\cal P}=\sum_{m=-\infty}^{\infty} \int_{0}^{\infty} \frac{d\omega }{2\pi} \, \hbar  \omega \, (n(\omega, T)-n(\omega,T_0)) \left(1- |S_m(\omega)|^2\right).
\end{equation}
We have expressed the radiation in terms of the Bose-Einstein distribution number $n(\omega,T)$. Clearly the net flux is in a direction opposite to the temperature gradient. The relation between the thermal emission and the absorptivity $1- |S|^2$, characterized by the deviation of the scattering matrix from unitarity, is \emph{Kirchhoff's law}~\cite{Beenakker98,Kruger11}. In the black-body limit, the object perfectly absorbs an incoming wave and does not reflect back, leading to the vanishing of the scattering matrix $S$. This is possible only if the dielectric function slightly deviates from 1 (otherwise, it leads to a finite scattering amplitude) with $\im \epsilon\ll 1$. While an infinite medium can be a perfect absorber at all frequencies and wave numbers, a compact object can act as a black body only in certain frequency regimes. At high temperatures the thermal radiation is dominated by large frequencies so we can assume $\im \epsilon\, \omega R \gg 1$.
Within these limits, it can be shown that the scattering matrix is almost unitary for $|m|> \omega R$ while it is approximately zero when $|m| < \omega R$. Therefore, the sum over $m$ at a fixed $\omega$ gives a factor of $2\omega R$ proportional to the circumference of the disk in harmony with the black-body radiation and Stefan-Boltzmann law~\cite{Boltzmann1884}.

In the following sections, we apply the techniques that we have developed here to rotating objects.

\subsection{{Field fluctuations for moving objects}}\label{Sec: Scalar corr fn Moving Object}
We first devise a Lagrangian from which Eq.~(\ref{scalar field equation}) follows for a static object, and then, with the guidance of Lorentz invariance, generalize it to a moving object. Schematically, the Lagrangian can be written as\footnote{The response function may be non-local in time; the Lagrangian merely serves as a guide to obtain the field equation.}
\begin{align}
  \mathcal L &= \frac{1}{2}\epsilon \, (\partial_t \Phi )^2-\frac{1}{2}(\nabla \Phi)^2 \nonumber \\
             &= \frac{1}{2}\left[(\partial_t \Phi )^2-(\nabla \Phi)^2\right] +\frac{1}{2}(\epsilon-1) \, (\partial_t \Phi )^2.
\end{align}
The second line breaks the Lagrangian into two parts: the first term is merely the free Lagrangian (in empty space) while the second term contributes only within the material, hence defining the interaction of the field with the object. In generalizing to moving objects, the free Lagrangian remains invariant. The interaction, however, should be defined with respect to the rest frame of the object. The latter is cast into a covariant form so that it reduces to the familiar expression in the rest frame
\begin{align}\label{Eq: Lagrangian for moving objects}
            {\cal L}= \frac{1}{2}(\partial_t \Phi )^2-\frac{1}{2}(\nabla \Phi)^2+\frac{1}{2}(\epsilon'-1) (U^\mu\partial_\mu \Phi)^2\,,
\end{align}
with $U$ being the four-velocity (or, three-velocity in 2+1 dimensional space-time) of the object. Note that $\Phi$ is scalar, i.e. $\Phi'(t',\bx')=\Phi(t,\bx)$ with the (un)primed coordinates defined in the (lab) comoving frame. Also the dielectric function $\epsilon'=\epsilon(\omega', \bx')$ is naturally defined in the comoving frame, and should be transformed to the coordinates in the lab frame.
Equation (\ref{Eq: Lagrangian for moving objects}) introduces a \emph{minimal }coupling between the object's motion and the scalar field in the background. For an object in uniform motion, this Lagrangian is obtained by an obvious Lorentz transformation. One might think that this equation should be further elaborated for an accelerating object. However, if the acceleration rate is small compared to the object's internal frequencies (plasma frequency, for example) the motion can be implemented by a local Lorentz transformation, hence Eq.~(\ref{Eq: Lagrangian for moving objects}). The field equation is deduced from the Lagrangian as
\begin{align}
   \left[\triangle-\partial_t^2- (\epsilon' -1)(U^\mu\partial_\mu)^2\right] \, \Phi(t,\bx)=0.\nonumber
\end{align}
This is the homogenous field equation in the presence of a moving object. We should also incorporate the coupling to random sources for applications of the Rytov formalism since the source is naturally defined in the comoving frame, a similar argument suggests a minimal coupling by adding $\Delta {\cal L}=- \varrho \, U^\mu\partial_\mu \Phi $ to the Lagrangian. The governing equation for the scalar field is then
\begin{align}\label{relativist Phi-rho equation}
  &-\left[\triangle-\partial_t^2- (\epsilon' -1)(U^\mu\partial_\mu)^2\right] \, \Phi(t,\bx)=U^\mu\partial_\mu \varrho(t,\bx),
\end{align}
which reduces to Eq.~(\ref{Eq: inhom field eqn static}) for an object at rest.
Here, we have {\it defined} $\varrho'(t',\bx')\equiv \varrho(t,\bx)$. Source fluctuations are distributed according to Eq.~(\ref{Eq: source fluc static}) but with respect to the comoving frame,
\begin{equation}\label{source fluc comov frame}
    \langle \varrho'_{\omega'}(\bx') {\varrho'}^*_{\omega'}(\by') \rangle =a(\omega')\im \epsilon(\omega', \bx') \, \delta(\bx'-\by'),
\end{equation}
with primed quantities defined in the moving frame. The two sets of coordinates are related via
\begin{equation}\label{Eq: coordinate transf}
\begin{cases}
  t'=t, & \\
  r'=r, & \\
  \phi'=\phi-\Omega t.
\end{cases}
\end{equation}

We shall limit ourselves only to objects moving at velocities small compared to the speed of light, in which case, $U\approx(1, \bv)$ with $\bv$ being the local velocity. Rotating at an angular frequency $\Omega$, $\bv={\bf\Omega} \times \bx$, Eq.~(\ref{relativist Phi-rho equation}) becomes
\begin{align}\label{Eq: nonrel eom inside}
  -\left[\triangle-\partial_t^2- (\epsilon' -1) (\partial_t + \Omega \partial_\phi)^2\right]\, \Phi(t,\bx)= (\partial_t +\Omega \partial_\phi)\varrho(t,\bx).
\end{align}
Let us expand the random source $\varrho(t,\bx)$ in the lab frame as
\begin{equation}
  \varrho(t,\bx)=\int \frac{d\omega}{2\pi} \, e^{-i\omega t} \varrho_{\omega}(\bx) = \sum_{m}\int \frac{d\omega}{2\pi} \, e^{-i\omega t+im\phi} \varrho_{\omega, m}(r).
\end{equation}
Similarly, we define $\varrho'_{\omega',m'}$ in the comoving frame with $\omega'$ and $m'$ being conjugate to the time and angular variables in the same frame. The coordinate transformations in Eq.~(\ref{Eq: coordinate transf}) along with the definition $\varrho'(t',\bx')\equiv \varrho(t,\bx)$ yield $\varrho_{\omega,m}(r)= \varrho'_{\omega-\Omega m, m }(r)$. Therefore fluctuations in the comoving frame, Eq.~(\ref{source fluc comov frame}), translate to
\begin{equation}\label{rho-rho moving}
     \langle \varrho_{\omega,m}(r) {\varrho}^*_{\omega,m}(\xi) \rangle =a(\omega- \Omega m)\im \epsilon(\omega-\Omega m, r) \, \frac{r^{-1}\delta(r-\xi)}{2\pi} ,
\end{equation}
in the lab frame. This equation is indeed similar to source fluctuations in a static object with $\omega$ being replaced by $\omega-\Omega m$. In other words, zero-point fluctuations in the object are centered at a frequency shifted from that of the vacuum.

Having formulated field equations and their corresponding source fluctuations, we compute correlation functions in the next section.

\subsubsection{Field correlations}\label{Sec: field correlations}
Similar to Sec.~\ref{Field Fluctuations for Static Objects}, we compute the field correlation functions separately for source fluctuations outside and inside the object.
The treatment of the vacuum (outside) fluctuation is entirely identical to the case of a static object, Eq.~(\ref{static out fluc}), while the scattering matrix is generally different when rotating.

For inside source fluctuations, the argument should be modified slightly. Let us define the (new) functions $f$ as solutions to the wave equation inside the object
\begin{equation}
\left[\triangle-\partial_t^2 - (\epsilon' -1)(\partial_t + \Omega \partial_\phi)^2\right]\, e^{-i\omega t} e^{i m \phi}f_{\omega,m}(r) = 0.
\end{equation}
The Green's function for one point inside and the other outside the object takes a similar form to the static case \begin{align}
  G(\omega, \bx,\by)
  &=  \sum_{m=-\infty}^{\infty} \frac{i}{8} \, f_{\omega,m}(r) e^{i m\phi} \, H_m^{(1)}(\omega \xi) e^{-im\psi}, \hskip .3in r<R<\xi,
  \end{align}
with the function $f$ satisfying continuity relations similar to Eq.~(\ref{continuity}) with $S_{m}(\omega)$ replaced by $S_{-m}(\omega)$.\footnote{With time reversal invariance, the Green's function, $G(\omega, \bx,\by)$, is symmetric in its spatial arguments,
\begin{equation*}
  G(\omega, \bx,\by)=  G(\omega, \by,\bx).
\end{equation*}
For a rotating object, time reversal is no longer a symmetry; however, time reversal followed by reversing the angular velocity forms a symmetry which yields
\begin{equation*}
  G(\omega, r,\phi,\xi,\psi)=  G(\omega, \xi,-\psi, r,-\phi).
\end{equation*}
The negative sign carries through to the sign of the angular momentum $m$.
} The field correlation function is then related to source fluctuations as
\begin{align}
  \langle \Phi(\omega, \bx)\Phi^*(\omega, \by)\rangle_{\innf}
  =&  \frac{1}{64} \sum_{m=-\infty}^{\infty} (\omega-\Omega m)^2 \, a_{\inn}(\omega-\Omega m) \,
  H_m^{(1)}(\omega r) e^{im\phi} \overline{H_m^{(1)}(\omega \xi) e^{im\psi}} \times \nonumber \\
  & \,2\pi \int_{0}^{R} d\rho \, \rho \,f_{\omega, m}(\rho)\im \epsilon(\omega- \Omega m,\rho)\, \overline{f_{\omega, m}(\rho)},
\end{align}
where we have used Eq.~(\ref{rho-rho moving}).
As before, we can exploit the wave equation to convert the integral in the last equation to a boundary term. The correlation function can be then cast in terms of the scattering matrix as
\begin{equation}\label{moving in fluc}
  \langle \Phi(\omega, \bx)\Phi^*(\omega, \by)\rangle_{\innf}=\frac{1}{16} \sum_{m=-\infty}^{\infty}  a_{\inn}(\omega-\Omega m) \, \left(1-|S_m(\omega)|^2\right)   H_m^{(1)}(\omega r) e^{im\phi} \overline{H_m^{(1)}(\omega \xi) e^{im\psi}}.
\end{equation}
This equation is similar to the expression for a static object, Eq.~(\ref{static in fluc}), with the important difference that the distribution $a$ is a function of a shifted frequency defined from the point of view of the rotating frame.

\subsubsection{{Radiation, spontaneous emission and superradiance}}\label{Sec: Radiation}
In a Gaussian theory, two-point correlation functions define the complete structure of fluctuations, and can be used to compute force, torque or radiation. Specifically, the energy radiation per unit time is obtained by the integral of $\langle \partial_t \Phi \partial_r \Phi\rangle$ over a surface enclosing the object. For a rotating object, the correlation functions derived in the previous subsection yield
\begin{equation}\label{Eq: moving energy rad}
  {\cal P}=\sum_{m=-\infty}^{\infty} \int_{0}^{\infty} \frac{d\omega}{2\pi} \, \hbar \omega \, \left[n_{\inn}(\omega-\Omega m)-n_{\out}(\omega)\right] \left(1- |S_m(\omega)|^2\right).
\end{equation}
Similarly the torque, or the rate of angular momentum radiation, is given by integrating $\langle \partial_t \Phi \partial_\phi \Phi\rangle$ over the surface. We find an expression similar to Eq.~(\ref{Eq: moving energy rad}) by replacing $\hbar \omega$ by $\hbar m$,
\begin{equation}\label{Eq: moving L rad}
 M=\sum_{m=-\infty}^{\infty} \int_{0}^{\infty} \frac{d\omega}{2\pi} \, \hbar m\, \left[n_{\inn}(\omega-\Omega m)-n_{\out}(\omega)\right] \left(1- |S_m(\omega)|^2\right).
\end{equation}
The function $n_{\inn}$ is singular at $\omega=\Omega m$; however, at this frequency $\im \epsilon(\omega-\Omega m)=\im \epsilon(0)=0$ which results in no loss. Therefore, $1-|S|^2$ is zero at $\omega=\Omega m$ removing the singularity and rendering the above expressions well-defined. We stress that, at zero temperature, Eqs.~(\ref{Eq: moving energy rad}, \ref{Eq: moving L rad}) should be understood only to the leading order in $\Omega R/c$ as computing higher orders in this quantity requires a more careful treatment of the field equations in higher orders of velocity. At $T=T_0=0$, the sum over partial waves is restricted to positive $m$ where the leading contribution comes from $m=1$ while higher values of $m$ give the leading radiation at multipolarity $m$. At a finite temperature, the contribution due to higher partial values can be important, and even dominant, in which case they should be included. In the rest of this paper, summation over all partial waves should be understood in similar terms.
Nevertheless, the more general input-output formalism precisely gives Eqs.~(\ref{Eq: moving energy rad}, \ref{Eq: moving L rad}) without any approximations regarding the velocity of the rotating object \cite{Maghrebi13}, hence their validity goes beyond the analysis provided here.

Let us consider the limit of zero temperature so that thermal radiation can be neglected. In this limit, $n(\omega)=\, -\Theta(-\omega)$, that is the distribution function vanishes for positive frequency but becomes 1 for negative frequencies. This distribution defines a \emph{vacuum }state in which all positive-energy states are empty, and, figuratively, all negative energy states are occupied. Now the distribution function pertaining to inside fluctuations is defined with respect to a frequency shifted by a multiple of rotation frequency and thus can find negative values even when $\omega$ is positive. The difference of the Bose-Einstein distributions contributes in a frequency window of $[0,\Omega m]$. Therefore, even at zero temperature, a rotating object emits photons and loses energy; the number of photons emitted at frequency $\omega(>0)$ and partial wave $m$ is given by
\begin{equation}\label{photon radiation}
  {\cal N}_m(\omega)\equiv \frac{d{ N}_m(\omega)}{d\omega dt}=\Theta(\Omega m-\omega)\left(|S_m(\omega)|^2-1\right).
\end{equation}
The corresponding radiated energy or angular momentum is obtained by integrating over photon number multiplied by $\hbar \omega$ or $\hbar m$ respectively.
It follows from Eq.~(\ref{photon radiation}) that a (physically acceptable) positive outflux of photons requires a \emph{super-unitary} scattering matrix, $|S_{m}(\omega)|>1$.
Indeed Zel'dovich argued that classical waves should amplify upon scattering from a rotating object exactly for frequencies in a range $0<\omega<\Omega m$, a phenomenon which is called {\it superradiance}~\cite{Zel'dovich71}. While spontaneous emission by a rotating object is a purely quantum effect, superradiance can be understood entirely within classical mechanics: A system is lossy if the imaginary part of its response function is positive (negative) for positive (negative) frequencies. For a rotating object, $\im \epsilon(\omega')$ has the same sign as $\omega'=\omega-\Omega m$, the frequency defined in the comoving frame; however, for (positive) $\omega$ smaller than $\Omega m$, the argument of the dielectric function is negative and thus the object amplifies the corresponding incident waves, hence superradiance. In fact, incoming waves in the superradiating regime extract energy from a rotating object and slow it down.

Superradiance and  spontaneous emission are intimately related. When the object is at rest, it absorbs energy by getting excited to a higher level, and de-excites by emitting a photon. For a rotating body, this picture breaks down, that is the object can emit a photon while being excited to a higher level: The energy of the emitted photon is $\hbar \omega>0$ in the lab frame; however, a rotating observer sees the same particle at a shifted frequency $\omega'=\omega-\Omega m$. In the superradiant regime where $\omega<\Omega m$, the frequency is negative in the comoving frame, hence the object has gained (positive) energy.
This gain should be interpreted as heat generated inside the body. The energy conservation still holds because the energy of the emitted photon as well as heat are extracted from the rotational energy of the object. This observation is also at the heart of the superradiance phenomenon when incoming waves are enhanced upon scattering from a rotating object.
The above argument shows that spontaneous emission conserves the energy and thus is (energetically) possible. In fact, as the object spontaneously emits photons (and heats up), it also slows down unless kept in steady motion by an external agent. In the context of general relativity, the Penrose process provides a similar mechanism to extract energy from a rotating black hole~\cite{Penrose69}, which also leads to spontaneous emission~\cite{Unruh74}.

We define  $E$ and $E'$ as the energy of the object in the lab frame and the rotating frame, respectively. The two are related by $E'=E-\Omega L$ where $L$ is the angular momentum of the object~\cite{LandauLifshitzMechanics}. Hence, the heat generated per unit time, ${\cal Q} \equiv{dE'}/{dt}$, is given by
\begin{equation}\label{Eq: heat}
  {\cal Q} \equiv\frac{dE'}{dt}=\frac{dE}{dt}-\Omega\frac{dL}{dt}=\Omega M-{\cal P}.
\end{equation}
In order to maintain a steady rotation, one should exert a constant torque $M$. The work done is equal to the radiated energy plus heat, $\Omega M = {\cal P} + {\cal Q}$. Note that the object loses energy to the environment, $dE/dt=-{\cal P}<0$, as well as angular momentum, $dL/dt=-M<0$. The rate of the energy gain in the object's rest frame can be obtained from Eqs. (\ref{Eq: moving energy rad}) and (\ref{Eq: moving L rad}) as
\begin{equation}
  {\cal Q}=\sum_{m=-\infty}^{\infty} \int_{0}^{\infty} \frac{d\omega}{2\pi} \, \hbar (\Omega m-\omega)\, {{\cal  N}_m(\omega)}.
\end{equation}
At zero temperature, the photon number production, Eq.~(\ref{photon radiation}), has nonzero support only for $0<\omega<\Omega m$ and thus the heat generation is manifestly positive. In brief, the object heats up while it loses energy ($E$ decreases) if not connected to an infinite thermal bath. This suggests that the heat capacity from the point of view of the lab frame is negative; however, thermodynamic quantities are well-defined in the comoving frame where the energy, $E'$, increases, hence the heat capacity is indeed positive.

We have argued that spontaneous emission is energetically possible, consistent with the energy conservation. This process also generates heat inside the object and photons in the environment, hence entropy is increasing. Notice that the line of argument can be reversed: A phenomenon which satisfies requirements of energy conservation and is thermodynamically favored due to entropy production should occur. This observation completes the link between superradiance and spontaneous emission, see also Refs.~\cite{Zel'dovich71,Bekenstein98}.
In Sec.~\ref{Sec: Entropy}, we study the statistics of radiated photons in some detail. In particular, we compute the entropy generation due to the creation of photons.
\subsubsection{{Radiation: rotating disk}}
In this section, we study quantum radiation by a rotating disk of radius $R$ described by a spatially uniform but frequency-dependent dielectric function $\epsilon(\omega)$. We find solutions to the field equation inside and outside the object, and match them on the boundary to compute the scattering amplitude.
When linear velocities are small, Eq.~(\ref{Eq: nonrel eom inside}) for the field equation (with the source term in the RHS set to zero) yields
\begin{equation}
  \left[\triangle- \partial_t^2-(\epsilon'-1)\left(\partial_t+\Omega \partial_\phi\right)^2\right]\Phi(t,\bx)=0.
\end{equation}
A solution characterized by frequency $\omega$ and the angular momentum $m$, i.e. of the form $\Phi= f(r)e^{-i \omega t} e^{i m \phi}$, casts this equation to
\begin{equation}\label{nonrel inside sln}
  \left[\frac{1}{r}\partial_r r \partial_r -\frac{m^2}{r^2}+ \tilde\omega_m^2\right]f(r)=0.
\end{equation}
Here, we have defined a new $m$-dependent (possibly complex) frequency $\tilde\omega_m$ as
\begin{equation}\label{nonrel inside freq}
\tilde\omega_m^2=(\epsilon'-1)\left(\omega-\Omega m\right)^2+ \omega^2,
\end{equation}
which is a constant for a fixed $\omega$ and $m$, and position-independent $\epsilon'=\epsilon(\omega')=\epsilon(\omega-\Omega m)$. Therefore, the equation that governs the field dynamics inside the object is a Helmholtz equation whose regular solutions are Bessel-$J$ functions, with the frequency replaced by $\tilde\omega_m$. Note that both the order and the argument of the Bessel functions depend on $m$, the latter through $\tilde\omega_m$. We define a scattering ansatz as
\begin{equation}
  \Phi(\omega,\bx)=
\begin{cases}
 V_m(\omega) \, J_m(\tilde\omega_m r) e^{i m \phi}, & r<R, \\
H^{(2)}_m(\omega r)e^{i m\phi}+ S_m(\omega)H^{(1)}_m(\omega r)e^{i m\phi},& r>R,
\end{cases}
\end{equation}
with the outside solutions being a linear combination of incoming and (with the scattering matrix as the amplitude) outgoing waves. The scattering matrix can be easily obtained by matching boundary conditions,
\begin{equation}\label{nonrel scatt mat}
   S_m(\omega)= -\frac{\partial_R J_m(\tilde \omega_m R) H^{(2)}_m(\omega R)-J_m(\tilde \omega_m R) \partial_R H^{(2)}_m(\omega R)}{\partial_R J_m(\tilde \omega_m R) H^{(1)}_m(\omega R)-J_m(\tilde \omega_m R) \partial_R H^{(1)}_m(\omega R)}.
\end{equation}
When $\epsilon$ is real, i.e. for a loss-less material, the denominator is merely the complex conjugate of the numerator, and the scattering is unitary. Conversely, if $\epsilon$ has an imaginary part the scattering matrix is non-unitary. For a lossy object at rest, $\im \tilde\omega_m =|\omega| \im \sqrt\epsilon>0$ (for positive frequency) and $|S|^2<1$. For a spinning object, $\im \tilde \omega_m \propto \im \epsilon'\propto \sgn(\omega-\Omega m)$, hence the scattering matrix is sub-unitary for $\omega>\Omega m$ but super-unitary, $|S|^2>1$, in the superradiating range $\omega<\Omega m$.

One can now compute the radiation from the $S$-matrix. Assuming that the object's linear velocity is small, the radiation is strongest at frequencies comparable to $\Omega$, thus the first partial wave $m=1$ suffices, and the Bessel-$J$ functions can be expanded.
The scattering matrix deviates from unitarity by (restoring units of $c$), as
\begin{equation}
  |S_1(\omega)|^2-1 \approx -\frac{\pi}{8}\frac{\omega^2 (\omega-\Omega)^2R^4}{c^4} \im \epsilon(\omega-\Omega).
\end{equation}
This expression is manifestly negative for $\omega>\Omega$ but positive when $\omega<\Omega$ for any causal $\epsilon$.
One can then compute various quantities of interest such as torque, heat generation, and radiation. In particular, energy radiation per unit time is given by Eq.~(\ref{Eq: moving energy rad}) as
\begin{align}
{\cal P}
&\approx \frac{\hbar R^4}{16 c^4}\int_{0}^{\Omega} d\omega \,\omega^3 (\omega-\Omega)^2|\im \epsilon(\omega-\Omega)|.
\end{align}
For a specific dielectric function, the radiation can be computed explicitly.

\subsection{Higher dimensions, non-scalar field theories and Trace formulas}
The above results can be readily generalized to higher dimensions. For a cylinder extended along the third dimension, quantum radiation is given by
\begin{equation}\label{Eq: radiation cylinder 3D}
  {\cal P}=\int_{0}^{\infty} \frac{d\omega }{2\pi} \, \hbar\omega \, \sum_{m=-\infty}^{\infty} \int_{-\omega}^{\omega}\frac{L dk_z}{2\pi}\left[n_{\inn}\, (\omega-\Omega m)-n_{\out}(\omega)\right] \left(1- |S_{mk_z}(\omega)|^2\right),
\end{equation}
where $L$ is the length of the cylinder, and $k_z$ is the wavevector along the $z$ direction. Note that $|k_z|$ is bounded by $\omega$ (we have set $c=1$) corresponding to propagating waves as opposed to evanescent waves which affect short distances from the cylinder but do not contribute to the radiation at infinity.

If the rotating object is not translationally symmetric in the $z$ direction (while rotationally symmetric), the scattering matrix is no longer diagonal in $k_z$ leading to a more complicated analog of Eq.~(\ref{Eq: radiation cylinder 3D}). Nevertheless, the $S$-matrix can always be diagonalized in some basis. Indeed one can write a general Trace formula for the quantum radiation which is independent of a particular basis,
\begin{equation}\label{Eq: basis free P}
  {\cal P}=\int_{0}^{\infty} \frac{d\omega }{2\pi} \, \hbar\omega \, \mbox{Tr} \left[\left(n_{\inn}\, (\omega-\Omega \hat l_z)-n_{\out}(\omega) \right)\left(1- {\mathbb S}{\mathbb S}^\dagger\right)\right],
\end{equation}
where we trace over all the propagating modes. In this equation, $\hat l_z=\frac{1}{i}\frac{\partial}{\partial \phi}$ is the angular momentum operator (in units of $\hbar$) projecting out the rotational index $m$. The scattering matrix $\mathbb S$ is written in a general basis-free notation. Equation~(\ref{Eq: basis free P}) is not specific to scalar fields or translationally symmetric objects but also holds for arbitrary shapes (though rotationally symmetric) and electromagnetism---the latter requires tracing over polarizations too. We present a general derivation of Eq.~(\ref{Eq: basis free P}) in Sec.~\ref{Sec: EM} in the context of electrodynamics.

\subsection{Photon statistics and entropy generation}\label{Sec: Entropy}
Heretofore, we have studied in some detail an object out of thermal or dynamic equilibrium with the environment, where it is shown that the object emits photons. In this section, we turn to a different aspect of this problem, namely the statistics of radiated photons.

We first note that the field correlation function receives contributions from photons as well as zero-point and (at finite temperature) thermal fluctuations, and can be broken up as
\begin{equation}\label{Eq: rad + non-rad}
  \langle \Phi \Phi\rangle =  \langle \Phi \Phi\rangle_{\rm non-rad}+\langle \Phi \Phi\rangle_{\rm rad} .
\end{equation}
The first term on the RHS is the non-radiative term,
\begin{equation}\label{Eq: non-radiation c.f.}
  \langle \Phi(\omega, \bx)\Phi^*(\omega, \by)\rangle_{\rm non-rad}=\hbar \coth\left(\frac{\hbar \omega}{2 k_B T}\right)\im G(\omega,\bx,\by).
\end{equation}
This expression is purely real and thus does not contribute to the radiation. Equation~(\ref{Eq: non-radiation c.f.}) is similar to the fluctuation-dissipation relation in equilibrium; cf. Eq.~(\ref{Eq: eq FDT1}). However, out of equilibrium, the total correlation function receives another contribution which cannot be written in the above form.
For a disk rotating at a rate $\Omega$, possibly at a finite temperature $T$, the radiation term can be deduced from the total correlation function (see Sec.~\ref{Sec: field correlations}), and using the above definition we find
\begin{equation}\label{Eq: radiation c.f.}
  \langle \Phi(\omega, \bx)\Phi^*(\omega, \by)\rangle_{\rm rad}=\frac{\hbar}{8} \sum_{m=-\infty}^{\infty} n(\omega-\Omega m,T) \left(1-|S_m(\omega)|^2\right)   H_m^{(1)}(\omega r) e^{im\phi} \overline{H_m^{(1)}(\omega \xi) e^{im\psi}}.
\end{equation}
This term is entirely composed of outgoing fields as expected.
In the remainder of this section, we focus on the ensemble of radiated photons.

Radiation can be quantified by the photon current, or the number of photons  radiated per unit time. Different frequencies and partial waves are statistically independent, thus we consider the current of a single mode of frequency $\omega$ and angular momentum $m$,
\begin{equation}\label{Eq: I-m}
  {\cal I}_{\omega,m}=\frac{2\pi r }{i}\left[\Phi_m^*(\omega,\bx) \, \partial_r \Phi_m(\omega,\bx)-\mbox{c.c.}\right],
\end{equation}
where the field is expanded over partial waves as $\Phi(\omega,\bx)=\sum_m \Phi_m(\omega,\bx)$.
When averaged over the radiation field, this expression reproduces Eq.~(\ref{photon radiation}) for a rotating object at $T=0$, or, more generally at a finite $T$,
\begin{equation}
{\cal N}_{m}(\omega)= \langle {\cal I}_{\omega,m} \rangle=n(\omega-\Omega m,T) \left(1-|S_m(\omega)|^2\right).
\end{equation}
We are interested in higher statistical moments for which we have to compute the corresponding correlation functions of currents. Since fluctuations are Gaussian-distributed, current correlation functions can be reduced to a product of two-point functions of fields according to Wick's theorem.

We compute the fluctuations of the current at the radiation zone far away from the object (keeping in mind that the radiation field in Eq.~(\ref{Eq: radiation c.f.}) is strictly outgoing), in which limit the radial derivative acting on $\Phi$ gives a factor of $i\omega/c$. Therefore, far from the object, the current defined in Eq.~(\ref{Eq: I-m}) can be cast as
\begin{align}\label{Eq: I-m simple}
  {\cal I}_{\omega, m}
   = \lim_{r\to \infty} \frac{4\pi\omega r}{c} \,\,\Phi_m^*(\omega,\bx) \, \Phi_m(\omega,\bx);
\end{align}
this expression is useful in evaluating $n$-point correlation functions.

We can also define the probability distribution function $P(n)$ with $n$ being the number of photons per mode emitted in a time duration $t$. We drop the subscript indices as the statistics can be computed independently for each mode.
The probability distribution is related to current correlators by the Glauber-Kelley-Kleiner formula~\cite{Glauber63,Kelley64},
\begin{equation}\label{Eq: Prob n photons}
  P(n)=\frac{1}{ n!}\,\langle I^n \, e^{-I}\rangle_{\rm rad}.
\end{equation}
We introduce a generating function $F(\eta)$
\begin{equation}\label{Eq: generating function}
  e^{F(\eta)}=\langle e^{\eta I}\rangle,
\end{equation}
which allows to compute the probability distribution from the generating function as
\begin{equation}\label{Eq: P from generating function}
  P(n)=\lim_{\eta\to-1}\frac{1}{n!}\frac{d^n}{d\eta^n}e^{F(\eta)}.
\end{equation}
Taylor-expanding $F$ in $\eta$ generates the cumulants of factorial moments as \cite{Beenakker1998photon}
\begin{equation}
  F(\eta)=\sum_{p=1}^{\infty} \frac{\kappa_p \eta^p}{p\,!}.
\end{equation}

For a single object discussed above, the current $I$ in Eq.~(\ref{Eq: I-m simple}) is a bilinear term in the field $\Phi$ and its conjugate. Diagrammatically, we can represent $I$ as a vertex with an incoming and an outgoing line corresponding to $\Phi^*$ and $\Phi$ respectively. From Eq.~(\ref{Eq: generating function}), it is then clear that the cumulants $\kappa_p$ are given by
\begin{equation} \label{Eq: cumulant}
  \kappa_p= \langle I^p \rangle_c,
\end{equation}
with the subscript $c$ indicating that the \emph{connected }component of the $p$-point function should be computed.
A little thought shows that the connected correlation function in the last equation yields
\begin{equation}\label{Eq: kappa-p}
  \kappa_p= (p-1)! \, {\cal N}^p,
\end{equation}
with ${\cal N}=\langle I\rangle$ being the average current per mode. The generating function is then
\begin{equation}\label{Eq: strong Kirchhoff}
  F(\eta)=-\log (1-\eta {\cal N}), \qquad \mbox{or,} \qquad e^{F(\eta)}=\frac{1}{1-\eta{\cal N}}.
\end{equation}
These equations indicate that the counting distribution, $P(n)$, is solely determined from the mean value of the radiation. This strong version of Kirchhoff's law is due to Bekenstein and Schiffer~\cite{Bekenstein94}; see also Ref.~\cite{Beenakker1998photon}.
$F$ can also be interpreted as a one-loop effective action in a background defined by $\eta I$. Adopting this point of view, Eqs.~(\ref{Eq: cumulant}) and (\ref{Eq: strong Kirchhoff}) follow immediately.
The probability distribution is easily deduced from Eqs.~(\ref{Eq: P from generating function}) and (\ref{Eq: strong Kirchhoff}) as
\begin{equation}
  P(n)= \frac{{\cal N}^n}{({\cal N}+1)^{n+1}}.
\end{equation}
This equation completely determines photon number statistics~\cite{Bekenstein94}. In particular, it yields the average and the variance of the number of radiated photons per mode
per unit time as
\begin{align}
  \langle {\cal I} \rangle ={\cal N},\qquad \mbox{Var }{\cal I}\equiv\left\langle \left({\cal I}-\langle {\cal I}\rangle\right)^2\right\rangle = {\cal N} ({\cal N}+1).
\end{align}
Having the full statistics, we can compute the entropy of radiated photons as
\begin{align}\label{Eq: entropy}
  \frac{S} {k_B}&=-\sum_{n=0}^{\infty} P(n)\log P(n)\nonumber \\
   &=({\cal N}+1)\log ({\cal N}+1)-{\cal N}\log{\cal N}.
\end{align}
In fact, this equation describes the entropy of a bosonic system out of equilibrium~\cite{LandauLifshitzS180}. If the occupation number $\cal N$ obeys the Bose-Einstein distribution, Eq.~(\ref{Eq: entropy}) indeed produces the entropy of a gas of thermal bosons.

Quantum or thermal radiation from a single object consists of photons across the whole spectrum. Therefore, we should sum over all frequencies and quantum numbers
\begin{equation*}
  \sum_{\omega,m}\quad \rightarrow \quad t\int\frac{d\omega}{2\pi}\sum_m\,\,,
\end{equation*}
where $t$ is the time interval under consideration. The entropy from Eq.~(\ref{Eq: entropy}) is then linearly increasing over time giving rise to a constant rate of entropy generation (restoring $\omega$ and $m$) as
\begin{equation}\label{Eq: entropy generation}
  {{\cal S}}\equiv \frac{dS}{dt}=k_B\sum_m\int_0^\infty\frac{d\omega}{2\pi} \,\,\left(({\cal N}_m(\omega)+1)\log ({\cal N}_m(\omega)+1)-{\cal N}_m(\omega)\log{\cal N}_m(\omega)\right).
\end{equation}
In the black-body limit (for a perfectly absorbing object at rest), we recover the entropy associated with Planckian radiation. For a finite-size object (comparable with thermal wavelength), the spectrum approaches that of the grey-body radiation where one should include the dependence on absorptivity $r\equiv 1-|S|^2$. Equation (\ref{Eq: entropy generation}) then depends on temperature, object's length scale, and material properties in a complicated way. Additionally, the object loses energy thus contributes negatively to entropy generation as ${\cal S}_{\rm object}=-{\cal P}/T$ with $\cal P$ being the (mean) power. The total entropy increase per mode is then
\begin{equation}\label{Eq: total entropy}
  \frac{{\cal S}_{\rm total}}{k_B}=\left(\frac{r}{e^x-1}+1\right)\log\left(\frac{r}{e^x-1}+1\right)-\frac{r}{e^x-1} \log\frac{r}{e^x-1} - \frac{xr}{e^x-1},
\end{equation}
where $x=\hbar \omega/k_B T$ and $r$ is the absorptivity of the corresponding mode. It can be shown that this expression is positive for all $0\le r \le 1$ as expected.

Equation~(\ref{Eq: entropy generation}) can be understood as the \emph{entanglement entropy } between the object and the environment consisting of radiated photons. In Ref.~\cite{Klich09}, Klich and Levitov suggested that the entanglement entropy can be obtained from the full quantum statistics, or the \emph{quantum noise}. Specifically, the entanglement entropy generation at a quantum point contact (allowing electrons to transport between two leads) in the presence of a DC voltage $V$ was found to be $\frac{dS}{dt}=-\frac{eV}{h} \left[D\log D+(1-D)\log(1-D)\right]$ with $D$ being the transmission \cite{CWJB}. This expression is completely determined by the fluctuation of the electric current thus providing a link between quantum noise and entanglement entropy \cite{Klich09}. Equation~({\ref{Eq: entropy generation}}) indeed gives the bosonic analog of the results in Ref.~\cite{Klich09}, where the two leads should be thought of as the object and the environment. With the above picture in mind, Eq.~(\ref{Eq: total entropy}) finds a new interpretation: While the thermodynamic entropy of the object---the last term in Eq.~(\ref{Eq: total entropy})---decreases as the object loses energy, the sum of the entanglement entropy and the thermodynamic entropy always increases, indicating that the former indeed should be interpreted as entropy.

We are mainly interested in a rotating object at zero temperature with the radiation given by Eq.~(\ref{photon radiation}). Defining $\sigma\equiv|S|^2$, the entropy generation due to radiation from a rotating object is given by (with ${\cal N}_m(\omega)=\sigma_m(\omega)-1$)
\begin{equation}\label{Eq: entropy generation 2}
  {\cal S}=k_B\sum_{m=1}^{\infty} \int_0^{\Omega m}\frac{d\omega}{2\pi} \left[\sigma_m(\omega)\log \sigma_m(\omega)-(\sigma_m(\omega)-1)\log(\sigma_m(\omega)-1)\right].
\end{equation}
Similar to thermal radiation, there is another contribution to entropy due to the object itself. In this case, however, the latter is also increasing in time since the object heats up. Hence, as we have argued in Sec.~\ref{Sec: Radiation}, a rotating object tends to emit radiation for purely thermodynamic reasons.

Before concluding this section, we note that Eq.~(\ref{Eq: entropy generation 2}) can also be written as a Trace formula similar to the expression (\ref{Eq: basis free P}) for the energy radiation which should be valid in higher dimensions and other field theories including electrodynamics.

\subsection{Diffusion equation for rotation}\label{Sec. Diffusion}
We now examine the angular fluctuations of a spinning object as the result of the back-reaction force due to the radiation. Specifically, we find the probability distribution as a function of the angular velocity for a macroscopic object spinning freely or under a constant torque. Our discussion here applies to both zero temperature, which is dominated by zero-point quantum fluctuations, as well as to finite temperature.

We first consider an object freely rotating at an angular frequency $\Omega_0$. The radiation by the object carries away angular momentum parallel to the axis of rotation resulting in a decrease in angular velocity. We shall assume that the time duration under consideration, $t$, is much longer that $1/\Omega_0$, such that the radiated photons have definite frequencies. For simplicity, we first take this time sufficiently small such that the angular velocity does not change significantly. The frictional torque is obtained from the radiation current by $\hbar m {\cal I}$ summed (integrated) over all quantum numbers, so the average change in angular velocity is
\begin{equation}\label{Eq: M1}
    I\left(\Omega_0-\overline{\Omega(t)}\right)= \hbar t \sum_m \int_0^\infty\frac{d\omega}{2\pi}\,m {{\cal N}}_m(\omega)\equiv \hbar t \, \bar M(\Omega_0),
\end{equation}
where $I$ is the moment of inertia around the rotation axis, and $\bar M=M/\hbar$, the torque in units of $\hbar$, can be read off from Eq.~(\ref{Eq: moving L rad}) with $n_{\out}=0$.
Note that the dependence of the torque on the angular velocity, $\Omega_0$, is made explicit.
The variance of the angular momentum can be obtained from the corresponding variance of the current as
\begin{equation}\label{Eq: M2}
  \mbox{Var }I \Omega(t)=\hbar^2 t \sum_{m} \int_0^\infty\frac{d\omega}{2\pi}\,m^2 {{\cal N}}_m(\omega)\left({{\cal N}}_m(\omega)+1\right)\equiv \hbar^2 t \, \bar M_2(\Omega_0),
\end{equation}
where $\bar M_2$ is defined for future reference.
Exploiting the methods of the previous section, higher moments can be readily computed. For long times, however, the central limit theorem guarantees that the statistics is entirely determined by the mean and the variance of the distribution provided that the radiated photons are statistically independent. An extension of this theorem due to Lyapunov gives the statistical distribution even for long times when the initial angular velocity has changed significantly. The Lyapunov central limit theorem requires the random variables to be statistically independent but not necessarily identically distributed. With this assumption, the average of the random variables converges to a normal distribution with a mean value given by the sum of each variable's mean and a variance as the sum of all the variances~\cite{billingsley}. Equations (\ref{Eq: M1}) and (\ref{Eq: M2}) then take the form
\begin{align}\label{Eq: I Omega First Moment}
   I\left(\Omega_0-\overline{\Omega(t)}\right)&=\hbar \int^t dt' \,\bar M(\Omega(t')), \nonumber \\
  \mbox{Var }I \Omega(t)&= \hbar^2\int^t dt'\, {\bar M}_2(\Omega(t')),
\end{align}
which describe the deterministic decrease in the angular velocity as well as its uncertainty. Notice that the integrand in the above equation depends on the instantaneous value of the angular velocity. We stress that the above discussion is based on the adiabaticity of motion, namely, the rate at which the angular velocity $\Omega(t)$ changes is taken to be much smaller that $\Omega(t)$ itself.

The rotating object undergoes a stochastic motion due to the inherent quantum (and, at finite temperature, also thermal) fluctuations. Equivalently, the equation of motion can be cast into a Langevin equation subject to noise as
\begin{equation}\label{Eq: Langevin equation}
  I \dot\Omega(t) = -  \hbar \bar M(\Omega(t)) +\eta(t;\Omega(t)).
\end{equation}
The noise $\eta(t;\Omega(t))$ has zero mean
\begin{equation}
  \langle \eta(t;\Omega(t))\rangle=0,
\end{equation}
is independent at different times, and its covariance is
\begin{equation}
  \langle \eta(t;\Omega(t))\, \eta(t';\Omega(t'))\rangle=\hbar^2{ \bar M}_2(\Omega(t))\, \delta(t-t').
\end{equation}
The delta-function correlation in time implies that the radiated photons are not correlated over long times ($\gg 1/\Omega$).
One can easily check that Eq.~(\ref{Eq: I Omega First Moment}) follows directly from the Langevin equation (\ref{Eq: Langevin equation}). This equation is reminiscent of the Brownian motion for a particle due to its thermal motion where the angular velocity plays the role of the displacement. The Brownian motion is the prototype of the fluctuation-dissipation condition where the response function is related to the fluctuations in equilibrium. Equation (\ref{Eq: Langevin equation}) is rather distinct due to the fact that noise is evaluated out of equilibrium as the object rotates, hence the explicit dependence of the noise on $\Omega(t)$.\footnote{In the fluctuation-dissipation theorem, the noise is usually taken to be independent of the position and velocity of the particle within the linear response regime.} Nevertheless, we can deduce the distribution in angular velocity and its evolution just as one can find the probability distribution for a particle's position in a thermal bath. The Fokker-Planck equation offers a systematic derivation of the distribution function~\cite{kardar} which we denote by $\mathscr{P}(\Omega, t)$ making explicit the dependence on the angular velocity as a function of time. The master equation governing the probability distribution is
\begin{equation}
  \frac{\partial {\mathscr{P}}}{\partial t}+\frac{\partial }{\partial \Omega}\left[\frac{\hbar}{I} \bar M(\Omega) \mathscr{P} +\frac{\hbar^2}{I^2} \frac{\partial}{\partial \Omega}\left(\bar M_2(\Omega)\mathscr{P}\right)\right]=0.
\end{equation}
Notice that this equation reproduces the average and the variance in Eq.~(\ref{Eq: I Omega First Moment}) provided that the probability distribution is sharply peaked around the instantaneous average angular velocity. In other words, for an object starting to spin with a definite angular frequency, i.e. a delta function as $\delta(\Omega-\Omega_0)$, the time evolution of the probability distribution is, at long times, governed by a Gaussian function with the average and the variance given above.

In the presence of an external torque $M_0=\hbar \bar M(\Omega_0)$ which tries to keep the object at a constant angular velocity $\Omega_0$, the Fokker-Planck equation is modified as
\begin{equation}
  \frac{\partial {\mathscr{P}}}{\partial t}+\frac{\partial }{\partial \Omega}\left[\frac{\hbar}{I} \left(\bar M(\Omega) -\bar M(\Omega_0)\right)\mathscr{P} +\frac{\hbar^2}{I^2} \frac{\partial}{\partial \Omega}\left(\bar M_2(\Omega)\mathscr{P}\right)\right]=0.
\end{equation}
In the steady state where the probability distribution is constant in time, we find
\begin{equation}
  \frac{\hbar}{I} \left(\bar M(\Omega) -\bar M(\Omega_0)\right)\mathscr{P} +\frac{\hbar^2}{I^2} \frac{\partial}{\partial \Omega}\left({ \bar M}_2(\Omega)\mathscr{P}\right)=0.
\end{equation}
This equation can be solved exactly to obtain
\begin{equation}
{  \mathscr{P}}(\Omega)=\frac{C}{{ \bar M}_2(\Omega)}\exp\left[-\frac{I}{\hbar}\int_0^\Omega d\Omega' \, \frac{\bar M(\Omega')-\bar M(\Omega_0)}{\bar M_2(\Omega')}\right],
\end{equation}
where the normalization constant $C$ is determined by the condition $\int_0^\infty d\Omega\, \mathscr{P}(\Omega)=1$. For an object with a large moment of inertia, the distribution is sharply peaked near $\Omega_0$. In this case, the distribution function becomes a Gaussian in $\Omega$ as
\begin{equation}
  \mathscr{P}(\Omega)\approx \sqrt{2\pi}\,\,{\Delta} \exp{\left[-\frac{(\Omega-\Omega_0)^2}{2(\Delta \Omega)^2} \right]},
\end{equation}
with
\begin{equation}
  I \Delta \Omega=\sqrt{{\hbar }{I}\frac{ \bar M_2(\Omega_0)}{{\partial \bar M}/{\partial \Omega_0 }}}.
\end{equation}
This equation sets a quantum limit on how close to an eigenstate the angular velocity of a spinning object---driven by a constant torque---can be.
Being in a regime that the angular velocity is small compared to other frequency scales, one can assume that the scattering matrix only slightly deviates from unity such that ${\cal N}_{m}(\omega) \ll 1$. At zero temperature, the leading contribution in $\Omega R/c$ is given by $m=1$. One can see from Eqs.~(\ref{Eq: M1},\ref{Eq: M2}) that $\bar M_2 \approx \bar M$; hence, the last equation becomes
\begin{equation}\label{Eq: uncertainty}
  I \Delta \Omega=\sqrt{{\hbar }{I}\frac{ 1}{{\partial \log \bar M }/{\partial \Omega_0}}} \quad \propto \quad \sqrt{\hbar I \Omega_0}.
\end{equation}
The last relation follows from the fact that usually $\bar M(\Omega_0)$ is a power law in $\Omega_0$; see Sec.~\ref{Sec: Friction on rotating objects}. This means that the uncertainty in the angular momentum of a single object is not proportional to $\hbar$, but to the geometrical mean of $\hbar$ and the total angular momentum which is much greater than $\hbar$. Note that any infinitesimal dissipation gives rise to the uncertainty in Eq.~(\ref{Eq: uncertainty}) independent of the details and strength of loss.

\subsection{Test object: torque and tangential force}\label{Sec: test object scalar}
In this section, we consider a second, or a \emph{test}, object in the vicinity of the rotating body, and study the interaction between the two. Such interaction goes beyond the Casimir-Polder force~\cite{Casimir48-1} between two polarizable objects due to the presence of the radiation field near the test object. As we argue below, the latter is the dominant contribution to the force when the two objects are far apart. Let the objects be two disks of radii $R$ and $a$ separated by a distance $d$. We shall assume that $d\gg R, a$ and that the test object is at rest.
Our starting point is Eq.~(\ref{Eq: rad + non-rad}) where the correlation function is broken into non-radiative and radiative parts---the former is related to the imaginary part of the Green's function via Eq.~(\ref{Eq: non-radiation c.f.}), while the latter is given by Eq.~(\ref{Eq: radiation c.f.}). We shall assume that the two objects are separated far enough that a single-reflection computation of the radiation field off of the second object suffices.

It is useful to expand the radiation field around this object in order to compute its scattering. Hence, we introduce \emph{translation matrices} relating wave functions around two different origins, through
\begin{equation}
  {H^{(1)}_m(\omega r_1) e^{i m\phi_1}} =\sum_{n=-\infty}^{\infty}H^{(1)}_{n-m}(\omega d) {J_{n}(\omega r_2) e^{in\phi_2}},
\end{equation}
with $(r_2, \phi_2)$ being the coordinates with respect to the center of the second object. Upon scattering off the test object, the amplitude of the outgoing waves is given by the object's $S$-matrix designated as $\mathfrak S$,
 \begin{align}
   J_{m}(\omega r) e^{im\phi}=\frac{1}{2}\left(H^{(2)}_m(x)+H^{(1)}_m(x)\right) e^{im\phi}\quad \rightarrow  \quad \frac{1}{2}\left(H_{m}^{(2)}(\omega r) +\mathfrak S_m(\omega) H_m^{(1)} (\omega r)\right)e^{i m \phi}.
 \end{align}
 We can then write the scattering off of the second object as
 \begin{align}
   \langle \Phi(\omega, \bx)\Phi^*(\omega, \by)&\rangle_{\rm scat}
  =\frac{\hbar}{32} \sum_{m=1}^{\infty} n(\Omega m-\omega,T) \left(1-|S_m(\omega)|^2\right)   \times \nonumber\\
  &\left(\sum_{n=-\infty}^\infty H_{n-m}^{(1)}(\omega d)\left(H_{n}^{(2)}(\omega r) +\mathfrak S_{n}(\omega) H_{n}^{(1)} (\omega r)\right)e^{i n \phi}\right) \times \nonumber \\
  &\overline{\left(\sum_{p=-\infty}^\infty H_{p-m}^{(1)}(\omega d)\left(H_{p}^{(2)}(\omega \xi) +\mathfrak S_{p}(\omega) H_{p}^{(1)} (\omega \xi)\right)e^{i p \psi}\right)}.
\end{align}

Next we compute the torque exerted on the test object by the radiation field of the rotating body. Note that we have neglected the non-radiative term in Eq.~(\ref{Eq: rad + non-rad}) because it is given by the imaginary part of the Green's function which can be reduced to a potential energy. The two objects being symmetric, the energy function is indifferent to a rotation of the disk and thus makes no contribution to the torque. The radiation field, on the other hand, exerts a torque which is the integral of $\langle \partial_\phi \Phi\partial_r \Phi\rangle$ over a closed contour around the test object. Note that $\partial_\phi \to  i n$ and $\partial_r$ combine into the Wronskian of Bessel $H$ functions of the first and second kind. A little algebra yields, in the-first-reflection approximation,
\begin{equation}
  M_{2\leftarrow 1}=\frac{\hbar}{8\pi}\sum_{m>0,n}n \, \int_{0}^{\infty} {d\omega}\,  n(\omega-\Omega m,T) \left(1-|S_m(\omega)|^2\right)
\left|H^{(1)}_{n-m}(\omega d)\right|^2 \left(1-|\mathfrak S_{n}(\omega)|^2\right).
\end{equation}
The subscript indicates that the torque is exerted due to the radiation field of the first on the second object. As explained above, for a slowly rotating object at zero temperature, we may restrict to $m=1$. Further, $n=1$ is dominant at large separation. We then find the torque at $T=0$ as
\begin{equation}
  M_{2\leftarrow 1}=\frac{\hbar}{8\pi}\int_{0}^{\Omega } {d\omega}\, \left(|S_1(\omega)|^2-1\right)\left|H^{(1)}_{0}(\omega d)\right|^2 \left(1-|\mathfrak S_{1}(\omega)|^2\right).
\end{equation}
At close separations, one should include higher-order reflections. In the opposite extreme of large separations, $\Omega d /c\gg 1$, the torque falls off with distance as
\begin{equation}
M_{2\leftarrow 1}\sim\frac{\hbar c}{4\pi^2 d} \int_{0}^{\Omega } d\omega \frac{1}{\omega} \left(|S_1(\omega)|^2-1\right)\left(1-|\mathfrak S_{1}(\omega)|^2\right),
\end{equation}
where we have made the factos of $c$ explicit. Note that a non-vanishing torque requires the test object to be lossy, i.e. $|\mathfrak S_{1}(\omega)|<1$.

One can also compute the force exerted on the test object. Let the two objects be separated along the $x$ axis. Geometrically, they are symmetric with respect to the axis connecting them, nevertheless, a \emph{tangential }force arises in the perpendicular direction along the $y$ axis due to the radiation field. This force can be computed from the expectation value of the stress tensor
\begin{equation}
  T_{ij} = \partial_i \Phi\partial_j\Phi +\frac{1}{2}\delta_{ij}\left((\partial_t\Phi)^2-(\nabla\Phi)^2\right).
\end{equation}
To compute the force parallel to the $y$ axis, one  should integrate the expectation value of the stress tensor over a closed contour around the test object:
\begin{align}
  F_y&=r \int_{0}^{2\pi} d\phi \, \langle T_{ij}\rangle  \,\hat r_i \hat y_j \nonumber\\
     &= r \int_{0}^{2\pi} d\phi \left\langle\frac{1}{2} \sin \phi \left((\partial_t\Phi)^2+(\partial_r\Phi)^2-\frac{1}{r^2}(\partial_\phi\Phi)^2\right)+{\cos\phi}\frac{1}{r}\partial_r\Phi \partial_\phi\Phi\right\rangle.
\end{align}
Again, non-radiative terms do not contribute on the basis of symmetry. We can compute the tangential force explicitly; however, the algebra is rather long and the result is not very illuminating for our toy model of scalar fields. We postpone the discussion of the force to Sec.~\ref{Sec: test object} in the context of electromagnetism.

\section{Electrodynamics}\label{Sec: EM}
In this section, we generalize the methods and techniques developed in application to a scalar field to electromagnetism. The vector character of the latter complicates mathematical expressions, but the underlying concepts are identical to Sec.~\ref{Sec: Scalar field}, with the techniques straightforwardly extended to electrodynamics. We start from electromagnetic fluctuations and the corresponding correlation functions in the context of static objects, and generalize them to spinning objects. Throughout this section, we consider objects of arbitrary shape (rotationally symmetric in the case of spinning bodies) in a general basis of partial waves in three dimensions. We derive general trace formulas for the quantum and thermal radiation from a single object. We shall also explicitly keep the dependence on $c$.

\subsection{Static objects }
Quantum fluctuations of the electromagnetic field can be formulated in a number of ways. In a lossy medium such as a dielectric object, there are subtle complications requiring a careful treatment~\cite{Matloob95,Gruner96,Loudon97,Narayanaswamy08}. A convenient starting point for our purposes is the Rytov formalism~\cite{Rytov89} which relates quantum fluctuations of the fields to those of the sources and currents. For a dielectric object (with no magnetic response $\mu=1$), Maxwell equations in the presence of sources are
\begin{align}
  \begin{cases}
    \nabla \times \bE=i \frac{\omega}{c} \bB, & \\
    \nabla \times \bB=-i\,\epsilon(\omega) \, \frac{\omega}{c} \bE-i\frac{\omega }{c}\bK,  &
  \end{cases}
\end{align}
or equivalently
\begin{equation}\label{Eq: EM eqn}
 \left( \nabla \times \nabla \times -\, \frac{\omega^2}{c^2}\epsilon(\omega) \mathbb I \right)\bE= \frac{\omega^2}{c^2} \bK.
\end{equation}
Then, according to the Rytov formalism, source fluctuations are related to the imaginary part of the local dielectric function by
\begin{equation}\label{Eq: Source Fluc}
  \langle \bK(\omega,\bx) \otimes \bK^*(\omega,\by)\rangle =a(\omega) \im \epsilon (\omega, \bx) \delta(\bx-\by) \mathbb I,
\end{equation}
where the distributions $a$ and $n$ are defined as before. Current fluctuations are independent at different points (hence the delta function in space), and also independent for different vector components, hence the $3\times 3$ unit matrix $\mathbb I$. The corresponding fluctuations of the electromagnetic (EM) field can be described in terms of the sources from Eq.~(\ref{Eq: EM eqn}) via the EM Green's function, $\bE=\frac{\omega^2}{c^2}\int \mathbb G \bK$. We are mainly interested in the EM field fluctuations outside the object from which we can compute the quantum radiation. As we have discussed in the previous section, field fluctuations receive contributions both from the fluctuating sources within the object and from fluctuations (zero-point and at finite temperature, thermal) in the vacuum outside the object.
In the following, we first consider source fluctuations outside the object.

The dyadic EM Green's function is defined by
\begin{equation}
  \left(\nabla \times \nabla \times -  \frac{\omega^2}{c^2}\epsilon(\omega)\mathbb I \, \right)\,\mathbb G(\omega,\bx,\bz)= \mathbb I \, \delta (\bx-\bz).
\end{equation}
In an appropriate coordinate system $(\xi_1, \xi_2, \xi_3)$, the outgoing wave Green's function (in empty space) can be broken up along the coordinate $\xi_1$ as~\cite{Jackson98,Rahi09}
\begin{align}\label{Eq: G out-out0}
\mathbb G(\omega, \bx,\bz)=i
\begin{cases}
 \sum_\alpha  {\mathbf E}^{\out}_{\bar \alpha}(\omega,\bx)  \otimes {\mathbf E}^{\reg}_{\alpha}(\omega,\bz) & \xi_1(\bx)> \xi_1(\bz), \\
\sum_\alpha {\mathbf E}^{\reg}_{\alpha}(\omega,\bx) \otimes {\mathbf E}^{\out}_{\bar \alpha}(\omega,\bz) & \xi_1(\bx)< \xi_1(\bz),
\end{cases}
\end{align}
where $\bE^{\out(\inn)}$ is the outgoing (incoming) electric field normalized as
\begin{align}\label{Eq: identity}
  \frac{i}{2}\oint d{\mathbf \Sigma}\cdot \left[(\nabla \times \bE^{\out/\inn}_\alpha (\omega,\bz))\times \bE^{\out/\inn*}_{\beta}(\omega,\bz)+\bE^{\out / \inn}_\alpha (\omega,\bz)\times \nabla \times \bE^{\out/\inn*}_{\beta}(\omega,\bz)\right]&=\pm\delta_{\alpha \beta}\,, \\ \label{Eq: identity 2}
  \oint d{\mathbf \Sigma}\cdot \left[(\nabla \times \bE^{\out/\inn}_\alpha (\omega,\bz))\times \bE^{\inn/\out*}_{\beta}(\omega,\bz)+\bE^{\out / \inn}_\alpha (\omega,\bz)\times \nabla \times \bE^{\inn/\out*}_{\beta}(\omega,\bz)\right]&=0,
\end{align}
which can be derived by a vector form of the Green's theorem; see Appendix A. In other words, the vector field is normalized such that the corresponding current is unity up to a sign.
Also $\bE^{\reg}$ defines a solution to the EM field regular everywhere in space. The index $\alpha$ runs over partial waves, and $\bar\alpha$ indicates the partial wave which is related to $\alpha$ by time reversal.
In the presence of an external object, the free Green's function for both points outside the object should be modified to incorporate the scattering from the object (with $\xi_1(\bx)< \xi_1(\bz)$)
\begin{align}\label{Eq: G out-out1}
  \mathbb G(\omega,\bx,\bz)= \frac{i}{2}\sum_\alpha \left(\bE^{\inn}_\alpha(\omega,\bx)+S_\alpha(\omega) \bE^{\out}_\alpha(\omega,\bx)\right)\otimes \bE^{\out}_{\bar\alpha}(\omega,\bz) ,
\end{align}
where $S_\alpha(\omega)$ is the scattering matrix as a function of the frequency $\omega$ and partial wave $\alpha$. Note that we have assumed that the scattering matrix is diagonal in partial waves\footnote{We choose an appropriate coordinate system where Maxwell equations are separable, and take $\xi_1$ to be constant on the object's surface.}. In general, one should sum over all $\beta$ such that $S_{\beta\alpha}\ne 0$ with the rest of the derivation closely following the remainder of this section.  The incoming wave $\bE^{\inn}$ is normalized to ensure that the corresponding energy flux is $-\omega$. When the object is not present, $S_\alpha(\omega)=1$, and the last equation reduces to the free Green's function with $\bE^{\reg}_\alpha=(\bE^{\inn}_\alpha+ \bE^{\out}_\alpha)/2$. {We stress that the Green's function in Eq.~(\ref{Eq: G out-out1}) is defined with both points outside the object.}

The EM field correlation function due to the outside source fluctuations is then given by
\begin{align}
  \langle \bE (\omega,\bx) \otimes \bE^* (\omega,\by)\rangle_{\outf} &= \frac{\omega^4}{c^4} \int d\bz\langle \mathbb G(\omega, \bx,\bz) \bK (\omega,\bz)\otimes\mathbb G^*(\omega,\by,\bz) \bK^* (\omega,\bz)\rangle \nonumber \\
  &= a_{\out}(\omega) \frac{\omega^4}{c^4} \, \im \epsilon_D \int_{\out}  d\bz \, \mathbb G(\omega,\bx,\bz) \cdot \mathbb G^*(\omega,\by,\bz),
\end{align}
where the dot product indicates the contraction of the second subindex of the two dyadic functions. The last line in this equation is obtained according to Eq.~(\ref{Eq: Source Fluc}) where $a_{\out}$ corresponds to the distribution function at the environment temperature and $\epsilon_D$ is the ``dielectric function'' of the vacuum dust. The latter can be set to one only in the end as explained in the previous section: The integral over infinite space brings down a factor of 1/$\im \epsilon_D$~\cite{Kruger11}, while the integration over any finite region vanishes as we take the limit $\im \epsilon_D \to 0$. Therefore we can choose the domain of integration over $\bz$ such that $\xi_1(\bz)>\xi_1(\bx),\xi_1(\by)$. This allows us to use the partial wave expansion of the Green's function in Eq.~(\ref{Eq: G out-out1}) to find
\begin{align}
&\langle \bE (\omega,\bx) \otimes \bE^* (\omega,\by)\rangle_{\outf}=  \nonumber \\ &\frac{\omega^4}{4c^4}\, a_{\out}(\omega)\sum_{\alpha, \beta }\left(\bE^{\inn}_\alpha(\omega,\bx)+S_\alpha(\omega) \bE^{\out}_\alpha(\omega,\bx)\right)\otimes \left(\bE^{\inn*}_\beta(\omega,\by)+S_\beta^* \bE^{\out*}_\beta(\omega,\by)\right) \times \nonumber \\
&(\im \epsilon_D)\int_{\out} d\bz \, \bE_{\bar\alpha}^{\out}(\omega,\bz)\cdot \bE_{\bar\beta}^{\out*}(\omega,\bz).
\end{align}
Here and in subsequent parts, we frequently compute volume integrals similar to the last line of this equation, which can be cast as
\begin{align}
  &(\im \epsilon_D)\int_{\out} d\bz \, \bE_{\bar\alpha}^{\out}(\omega,\bz)\cdot \bE_{\bar\beta}^{\out*}(\omega,\bz)\nonumber\\
  =&\frac{1}{2i}\int_{\out} d\bz \, \left[\epsilon_D \,\bE_{\bar\alpha}^{\out}(\omega,\bz)\cdot \bE_{\bar\beta}^{\out*}(\omega,\bz)-\bE_{\bar\alpha}^{\out}(\omega,\bz)\cdot \epsilon_D^* \,\bE_{\bar\beta}^{\out*}(\omega,\bz)\right]\nonumber\\
  =&\frac{c^2}{2i\omega^2} \int_{\out} d\bz \, \left[(\nabla \times \nabla \times \,\bE_{\bar\alpha}^{\out}(\omega,\bz))\cdot \bE_{\bar\beta}^{\out*}(\omega,\bz)-\bE_{\bar\alpha}^{\out}(\omega,\bz)\cdot \nabla \times \nabla \times \,\bE_{\bar\beta}^{\out*}(\omega,\bz)\right]\nonumber,
\end{align}
where in the last line we have used the homogenous version of Eq.~(\ref{Eq: EM eqn}) with the RHS set to zero. The volume integration can be then recast as a surface integral with two boundaries, one at the infinity and another at a finite distance from the object. The infinitesimal imaginary part of the dielectric function guarantees that outgoing functions are exponentially decaying at large distances and thus the surface integral at infinity does not contribute. We then obtain
\begin{align}
&\im \epsilon_D\int_{\out} d\bz \, \bE_{\bar\alpha}^{\out}(\omega,\bz)\cdot \bE_{\bar\beta}^{\out*}(\omega,\bz) \nonumber \\
=&  \frac{i c^2}{2\omega^2} \oint d{\mathbf \Sigma}\cdot \left[(\nabla \times \bE^{\out}_{\bar\alpha }(\omega,\bz))\times \bE^{\out*}_{\bar\beta}(\omega,\bz)+\bE^{\out}_{\bar\alpha }(\omega,\bz)\times \nabla \times \bE^{\out*}_{\bar\beta}(\omega,\bz)\right].
\end{align}
Therefore, from Eq.~(\ref{Eq: identity}), the correlation function of the EM fields takes the form
\begin{align}\label{Eq: <EE> out static}
\langle \bE (\omega,\bx) \otimes \bE^* (\omega,\by)\rangle_{\outf}= &  a_{\out}(\omega)\frac{\omega^2 }{4c^2} \sum_{\alpha} \left(\bE^{\inn}_\alpha(\omega,\bx)+S_\alpha(\omega) \bE^{\out}_\alpha(\omega,\bx)\right)\otimes \nonumber \\ &\left(\bE^{\inn*}_\alpha(\omega,\by)+S^*_\alpha(\omega) \bE^{\out*}_\alpha(\omega,\by)\right).
\end{align}
The radiation due to the outside fluctuations can be computed by integrating over the Poynting vector, $\bS=c\, \bE \times \bB$, of the corresponding correlation function,
\begin{align}
 {\cal P}_{\outf}&= \int_{0}^{\infty} \frac{d\omega}{2\pi}\oint d{\mathbf \Sigma} \cdot \frac{ic^2}{ \omega} \langle (\nabla \times \bE)\times \bE^* +\bE \times \nabla \times \bE^* \rangle_{\outf} \nonumber \\
 &= \int_{0}^{\infty} \frac{d\omega }{2\pi}\, \sum_{\alpha}  (-1+ |S_\alpha(\omega)|^2) \,a_{\out}(\omega)\frac{i\omega}{4}\oint d{\mathbf \Sigma} \cdot \left[(\nabla \times \bE^{\out}_\alpha)\times \bE_\alpha^{\out*} +\bE^{\out}_\alpha \times \nabla \times \bE_\alpha^{\out *} \right] \nonumber \\
  &=    \frac{1}{4\pi}\int_{0}^{\infty} d\omega \, \omega \,a_{\out}(\omega)\, \sum_{\alpha} (-1+ |S_\alpha(\omega)|^2),
\end{align}
where we used Eqs.~(\ref{Eq: identity}) and (\ref{Eq: identity 2}).

The field correlation function induced by the inside fluctuations can be computed similarly. In this case, however, we need the Green's function with one point inside the object. Following an argument similar to the scalar case, we note that as the two points do not coincide, the Green's function satisfies a {\it homogeneous} equation inside with respect to the {\it smaller} coordinate while it satisfies the free EM equation outside the object in the {\it larger} coordinate. Hence, we can expand the Green's function as (with $\xi_1(\bz)<\xi_1(\bx)$)
\begin{equation} \label{Eq: Green fn in-out 0}
  \mathbb G(\omega, \bx,\bz)=  \frac{i}{2}\sum_{\alpha}  \left( A \, \bE^{\out}_{\bar\alpha}(\omega,\bx) +B \, \bE^{\inn}_{\bar\alpha}(\omega,\bx)\right)\otimes \bF_\alpha(\omega,\bz),
\end{equation}
where the prefactor ${i}/{2}$ is chosen for convenience, $A$ and $B$ are constants to be determined, and $\bF_\alpha$ is defined as the regular (at the origin) solution to the EM equation inside the object
\begin{equation}
\left(\nabla \times \nabla \times- \,  \frac{\omega^2}{c^2} \, \epsilon(\omega, \bx)\mathbb I\right) \bF_\alpha(\omega,\bx)=0.
\end{equation}
We can determine the coefficients $A$ and $B$ and the normalization of $\bF$ by matching the Green's functions approaching a point on the boundary from inside and outside the object
\begin{equation}
 \mathbb G(\omega, \bx,\by){\mid}_{\by\to \Sigma^{-}}=  \mathbb G(\omega, \bx,\by){\mid}_{\by \to \Sigma^+},  \hskip .2in \mbox{} \xi_1(\bx)>\xi_1(\by),
\end{equation}
where $\Sigma$ represents the boundary.
Comparing the two Green's functions given by Eqs. (\ref{Eq: G out-out1}) and (\ref{Eq: Green fn in-out 0}), we find ($A=1, B=0$)
\begin{align}
  \mathbb G(\omega, \bx,\bz)= \frac{i}{2}\sum_{\alpha}  \bE^{\out}_{\bar\alpha}(\omega,\bx)\otimes \bF_\alpha(\omega, \bz),
\end{align}
where $\bF_{\alpha}$ and the $S$-matrix element $S_{\alpha}$ are determined by the continuity of the Green's function which requires parallel components of electric and magnetic (the latter because $\mu=1$) fields to match at the boundary
\begin{align}\label{Eq: Boundary condition}
  \bF_\alpha(\omega, \bz)_\|&= \left(\bE^{\inn}_\alpha(\omega,\bz)+S_\alpha(\omega) \bE^{\out}_\alpha(\omega,\bz)\right)_{\|}, \nonumber\\
    \left(\nabla \times \bF_\alpha(\omega,\bz)\right)_\|&= \left(\nabla \times \bE^{\inn}_\alpha(\omega,\bz)+S_\alpha(\omega) \nabla \times \bE^{\out}_\alpha(\omega,\bz)\right)_{\|}.
\end{align}
The correlation function due to the inside fluctuations is then given by
\begin{align}
&\langle \bE (\omega,\bx) \otimes \bE^* (\omega,\by)\rangle_{\innf}\nonumber \\
=  &a_{\inn}(\omega)\,\frac{\omega^4}{4c^4} \sum_{\alpha, \beta }\bE^{\out}_{\bar\alpha}(\omega,\bx)\otimes \bE^{\out*}_{\bar\beta}(\omega,\by)  \int_{\inn} d\bz \,\bF_{\alpha}(\bz)\cdot \im \epsilon(\omega, \bz)\,\bF_{\beta}^{*}(\bz),
\end{align}
where $a$ is the distribution function defined at the object's temperature. Again exploiting the wave equation for $\bF$, the volume integral can be cast as a surface term
\begin{align}
  &\int_{\inn} d\bz \, \bF_{\alpha}(\bz)\cdot \im \epsilon(\omega, \bz)\,\bF_{\beta}^{*}(\omega,\bz) \nonumber \\
  =& \frac{c^2}{2i\omega^2} \oint d{\mathbf\Sigma}\cdot  \left[\left(\nabla \times \bF_\alpha (\omega,\bz)\right)\times \bF^{*}_{\beta}(\omega, \bz)+\bF_\alpha (\omega,\bz)\times \nabla \times \bF^{*}_{\beta}(\omega,\bz)\right].
\end{align}
The continuity equations can be used to evaluate the surface integral
\begin{equation}\label{Eq: Integral over F: static}
  \int_{\inn} d\bz \, \bF_{\alpha}(\omega,\bz)\cdot \im \epsilon(\omega, \bz)\,\bF_{\beta}^{*}(\omega,\bz) =  \frac{c^2}{\omega^2}\delta_{\alpha\beta} \left(1-|S_\alpha(\omega)|^2\right).
\end{equation}
The field correlation function then becomes\footnote{We have changed $\alpha\ \to \bar\alpha$; note that $|S_{\bar\alpha}|=|S_{\alpha}|$ due to time reversal symmetry.}
\begin{align}
\langle \bE (\omega,\bx) \otimes \bE^* (\omega,\by)\rangle_{\innf}=  & a_{\inn}(\omega)\frac{\omega^2}{4 c^2} \sum_{\alpha}\left(1-|S_\alpha(\omega)|^2\right) \bE^{\out}_{\alpha}(\omega,\bx)\otimes \bE^{\out*}_{\alpha}(\omega,\by).
\end{align}
The radiation power due to the inside fluctuations can be computed from the corresponding correlation function as
\begin{align}
 {\cal P}_{\innf}&= \frac{1}{4\pi}\int_{0}^{\infty} d\omega \, \omega \,a_{\inn}(\omega)\, \sum_{\alpha} (1- |S_\alpha(\omega)|^2).
\end{align}
The total radiation per unit time is given by
\begin{align}\label{Eq: EM rad}
  {\cal P}&= \frac{1}{4\pi}\int_{0}^{\infty} d\omega \, \omega \,\left(a_{\inn}(\omega)-a_{\out}(\omega)\right)\, \sum_{\alpha} \left(1- |S_\alpha(\omega)|^2\right) \nonumber \\
  & =\int_{0}^{\infty} \frac{d\omega}{2\pi} \, \hbar\omega \,\left(n(\omega, T)-n(\omega,T_0)\right)\, \sum_{\alpha} \left(1- |S_\alpha(\omega)|^2\right),
\end{align}
where in the last line the radiation is expressed in terms of the Bose-Einstein distribution function. In brief, we have derived the Kirchhoff's law in the context of electrodynamics~\cite{Beenakker98,Kruger11}, and the partial waves also include electromagnetic polarizations. Notice that Eq.~(\ref{Eq: EM rad}) is independent of the coordinate system and the shape of the object. In a general basis that the scattering matrix is not diagonal, the sum over $\alpha$ is replaced by a double sum over incoming and outgoing modes as
\begin{align}
  &\sum_{\alpha,\beta}\left(\delta_{\beta\alpha}-|S_{\beta\alpha}(\omega)|^2\right) \nonumber \\
  &=\,\mbox{Tr}\left(\mathbb I -\mathbb S^\dagger(\omega) \mathbb S(\omega)\right),
\end{align}
which is cast as a manifestly invariant (trace) formula in the last line.

\subsection{Moving objects}
For bodies in uniform motion,
the equations in the previous (sub)section are applied in the rest frame of the object and then transformed to describe the EM-field fluctuations in the appropriate laboratory frame. With all contributions of the field correlation functions in a single frame, one can then compute various physical quantities of interest, such as forces, or energy transfer from one object to another, or to the vacuum. For nonuniform motion, we assume that the same equations apply locally to the instantaneous rest frame of the body~\cite{Bladel76}. This assumption should be valid as long as the rate of acceleration is less than typical internal frequencies characterizing the object, which are normally quite large.
The EM wave equation for a moving medium can be inferred from a Lagrangian. A dielectric object is described by
\begin{align}\label{Eq: EM Lagrangian}
  \mathcal L
   = \frac{1}{2} \epsilon'\, \bE'^2 - \frac{1}{2} \bB'^2,
\end{align}
where $\bE'$ and $\bB'$ are the EM fields in the comoving frame related to the EM fields in the lab frame as
\begin{equation}\label{Eq: EM Lorentz trans}
  \bE'=\bE+{\frac{\mathbf v}{c}}\times \bB, \hskip .2in   \bB'=\bB-{\frac{\mathbf v}{c}}\times \bE,
\end{equation}
to the lowest order in velocity. Note that $\epsilon'=\epsilon (\omega',\bx')$ is the dielectric function defined in the moving frame similarly defined in Sec.~\ref{Sec: Scalar field}.
The Lagrangian can be cast as
\begin{equation}\nonumber
    \mathcal L
   = \mathcal L_0 +\frac{1}{2} (\epsilon'-1)\, \bE'^2,
\end{equation}
where $\mathcal L_0=\frac{1}{2} \, \bE'^2 - \frac{1}{2} \bB'^2$ is the free Lagrangian. Notice that $\mathcal L_0$ is invariant under the transformation in Eq.~(\ref{Eq: EM Lorentz trans}) to the first order in $v/c$, {\it i.e.} $\mathcal L_0=\frac{1}{2} \, \bE^2 - \frac{1}{2} \bB^2 +\mathcal O(v^2/c^2)$, while the second term is related to the EM field in the lab frame by Eq.~(\ref{Eq: EM Lorentz trans}).
The modified Maxwell equations are then obtained from the Lagrangian as
\begin{equation}\label{Eq: EM eqn moving}
  \left[\nabla \times \nabla \times -\frac{\omega^2}{c^2} \mathbb I -\frac{\omega^2}{c^2} \tilde{\mathbb D} (\epsilon'-1) \mathbb D  \right] \bE=0,
\end{equation}
where
\begin{align}
  \mathbb D= \mathbb I+\frac{1}{i\omega}{\mathbf v}\times \nabla \times\,, \hskip .2in
  \tilde{\mathbb D}= \mathbb I+\frac{1}{i\omega}\nabla \times{\mathbf v}\times.
\end{align}
The coupling with the (fluctuating) currents can be formulated by adding to the Lagrangian
\begin{equation}
  \Delta\mathcal L= \bK'\cdot\bE',
\end{equation}
where $\bK'$ is defined in the moving frame. This equation follows from the assumption that a local current density is coupled to the electric field in the instantaneous rest frame of the corresponding point in the moving object; see the discussion in Sec.~\ref{Sec: Scalar corr fn Moving Object}.
The inhomogeneous EM equation in the presence of random currents follows from the Lagrangian as
\begin{equation}
  \left[\nabla \times \nabla \times -\frac{\omega^2}{c^2} \mathbb I -\frac{\omega^2}{c^2} \tilde{\mathbb D} (\epsilon'-1) \mathbb D  \right] \bE=\frac{\omega^2}{c^2} \tilde{\mathbb D} \bK'.
\end{equation}

Again we should compute field correlation functions due to the outside and inside current fluctuations separately. The former can be easily deduced from Eq.~(\ref{Eq: <EE> out static}) simply by inserting the scattering matrix for a rotating object,
\begin{align}\label{Eq: <EE> out moving}
\langle \bE (\omega,\bx) \otimes \bE^* (\omega,\by)\rangle_{\outf}= &  a_{\out}(\omega)\frac{\omega^2 }{4c^2} \sum_{\alpha_m}\left(\bE^{\inn}_{\alpha_m}(\omega,\bx)+S_{\alpha_m}(\omega) \bE^{\out}_{\alpha_m}(\omega,\bx)\right)\otimes \nonumber \\
&\left(\bE^{\inn*}_{\alpha_m}(\omega,\by)+S_{\alpha_m}^*(\omega) \bE^{\out*}_{\alpha_m}(\omega,\by)\right),
\end{align}
where the partial-wave index $\alpha_m$ includes $m$, the eigenvalue of the angular momentum along the $z$-direction (in units of $\hbar$).

The inside fluctuations, on the other hand, are defined with respect to the rest frame of the object,
\begin{equation}
  \langle \bK'(\omega',\bx') \otimes {\bK'}^*(\omega', \by')\rangle = a_{\inn}(\omega') \im \epsilon (\omega', \bx') \delta(\bx'-\by')\mathbb I.
\end{equation}
Consider $\bK'_{\omega' m'}(t',\bx')$, a fluctuation of the current characterized by the angular momentum $m'$ and frequency $\omega'$ in the rotating frame; we do not make the dependence of the current $\bK$ on other quantum numbers explicit as its fluctuations depend only on $\omega$ and $m$ as it will become clear shortly. For the sake of notational convenience, we define $\bK(t,\bx)\equiv\bK'(t',\bx')$ which captures current fluctuations in the lab-frame coordinates. Note that the two sets of reference frame are related by Eq.~(\ref{Eq: coordinate transf}) and supplemented by $z=z'$ along the symmetry axis of the object. One can then see that the partial wave $m$ is invariant with respect to the reference frame while the frequency is shifted as
\begin{equation}\nonumber
    \omega'=\omega-\Omega m.
\end{equation}
This modifies the spectral density of source fluctuations simply by replacing the frequency in $\epsilon$ and $a$ by $\omega-\Omega m$. Therefore, the inside source fluctuations from the point of view of the lab--frame observer are given by
  \begin{align} \label{Eq: Source fluc moving}
    \langle  & \bK_m(\omega,\bx) \otimes \bK_m^*(\omega,\by) \rangle = a_T(\omega-\Omega m) \im \epsilon (\omega-\Omega m, r,z) \frac{\delta(r_\bx-r_\by)\delta(z_\bx-z_\by)}{2\pi r}\,\mathbb I.
  \end{align}
Henceforth, we shall use the same notation $\mathbb G$ for the Green's function in the presence of a moving object corresponding to Eq.~(\ref{Eq: EM eqn moving}).
The EM field correlation function is given by
\begin{align}
  \langle \bE (\omega,\bx) \otimes \bE^* (\omega,\by)\rangle_{\innf} &= \frac{\omega^4}{c^4} \int_{\inn} d\bz\,\langle \mathbb G(\omega,\bx,\bz) \tilde{\mathbb D}{\bK} (\omega,\bz)\cdot\mathbb G^*(\omega,\by,\bz) {\tilde{\mathbb D}}^* {\bK}^* (\omega,\bz)\rangle.
\end{align}
We can expand the Green's function similar to the previous section as
\begin{align}
  \mathbb G(\omega,\bx,\bz)= \frac{i}{2}\sum_{\alpha_m}  \bE^{\out}_{\bar \alpha_m}(\omega,\bx)\otimes \bF_{\alpha_m}(\omega,\bz),
\end{align}
where $\bF$ is a solution to the {\it modified} EM equation inside the dielectric object
\begin{equation}\label{rel em inside sln}
  \left[\nabla \times \nabla \times -\frac{\omega^2}{c^2} \mathbb I -\frac{\omega^2}{c^2} \tilde{\mathbb D} (\epsilon'-1) \mathbb D  \right] \bF=0,
\end{equation}
and satisfies boundary conditions similar to Eq.~(\ref{Eq: Boundary condition}), albeit with the scattering matrices for a rotating object\footnote{Note that the scattering matrix is given for $\bar\alpha$. This is because the Green's function in the presence of a moving object is no longer symmetric with respect to its spatial arguments but satisfies a rather different symmetry; see the discussion in Sec.~\ref{Sec: field correlations}. If Eq.~(\ref{Eq: G out-out1}) defines the Green's function with $\xi_1(\bx)< \xi_1(\bz)$, then
\begin{align*}
  \mathbb G(\omega,\bz,\bx)= \frac{i}{2}\sum_\alpha \bE^{\out}_{\bar\alpha}(\omega,\bz)\otimes\left(\bE^{\inn}_\alpha(\omega,\bx)+S_{\bar\alpha}(\omega) \bE^{\out}_\alpha(\omega,\bx)\right).
\end{align*}
}
\begin{align}\label{Eq: Boundary condition2}
  \bF_{\alpha_m}(\omega, \bz)_\|&= \left(\bE^{\inn}_{\alpha_{m}}(\omega,\bz)+S_{\bar\alpha_{m}}(\omega) \bE^{\out}_{\alpha_{m}}(\omega,\bz)\right)_{\|}, \nonumber\\
    \left(\nabla \times \bF_{\alpha_m}(\omega,\bz)\right)_\|&= \left(\nabla \times \bE^{\inn}_{\alpha_{m}}(\omega,\bz)+S_{\bar \alpha_{m}}(\omega) \nabla \times \bE^{\out}_{\alpha_{m}}(\omega,\bz)\right)_{\|}.
\end{align}
The correlation function of the EM fields is then given by
\begin{align}
\langle \bE (\omega,\bx) \otimes \bE^* (\omega,\by)\rangle_{\innf}=  \, &\frac{\omega^4}{4c^4} \sum_{\alpha_{m}, \beta_m } a_{\inn}(\omega-\Omega m)\,\bE^{\out}_{\bar\alpha_m}(\omega,\bx)\otimes \bE^{\out*}_{\bar\beta_m}(\omega,\by) \times \nonumber \\
&\int_{\inn} d\bz \, \mathbb D\,\bF_{\alpha_m}(\omega,\bz)\cdot \im \epsilon(\omega-\Omega m, \bz)\,{\mathbb D}^*\bF_{\beta_m}^{*}(\omega,\bz).
\end{align}
The volume integral can be computed similar to that of the previous subsection. We write the second line of the last equation as
\begin{align}
&\frac{1}{2i}\int_{\inn} d\bz \left[ (\epsilon'-1) \mathbb D\,\bF_{\alpha_m}(\omega,\bz)\cdot \,{\mathbb D}^*\bF_{\beta_m}^{*}(\omega,\bz)-\mathbb D\,\bF_{\alpha_m}(\omega,\bz)\cdot (\epsilon'^*-1)\,{\mathbb D}^*\bF_{\beta_m}^{*}(\omega,\bz)\right] \nonumber \\
=&\frac{1}{2i}\int_{\inn} d\bz \left[ \left(\tilde {\mathbb D}(\epsilon' -1)\mathbb D\,\bF_{\alpha_m}(\omega,\bz)\right)\cdot \bF_{\beta_m}^{*}(\omega,\bz)-\bF_{\alpha_m}(\omega,\bz)\cdot \tilde{\mathbb D}^*(\epsilon'^*-1)\,{\mathbb D}^*\bF_{\beta_m}^{*}(\omega,\bz)\right] \nonumber \\
=&\frac{c^2}{2i\omega^2}\int_{\inn} d\bz \left[ \left(\nabla\times\nabla \times\bF_{\alpha_m}(\omega,\bz)\right)\cdot \bF_{\beta_m}^{*}(\omega,\bz)-\bF_{\alpha_m}(\omega,\bz)\cdot \nabla\times\nabla \times\bF_{\beta_m}^{*}(\omega,\bz)\right] \nonumber \\
=&\frac{c^2}{2i\omega^2}\oint d{\mathbf\Sigma}\cdot \left[ \left(\nabla \times\bF_{\alpha_m}(\omega,\bz)\right)\times \bF_{\beta_m}^{*}(\omega,\bz)+\bF_{\alpha_m}(\omega,\bz)\times \nabla \times\bF_{\beta_m}^{*}(\omega,\bz)\right],
\end{align}
where in the step from the second to the third line, we have used Eq.~(\ref{Eq: EM eqn moving}).
Using the continuity relations, the last line gives
\begin{align}
  \int_{\inn} d\bz \, \mathbb D\,\bF_{\alpha_m}(\omega,\bz)\cdot \im \epsilon(\omega-\Omega m, \bz)\,{\mathbb D}^*\bF_{\beta_m}^{*}(\omega,\bz)= \frac{c^2}{\omega^2}\, \delta_{\alpha_m \beta_m} \left(1-|S_{\bar\alpha_{m}}|^2\right),
\end{align}
 which is the analog of Eq.~(\ref{Eq: Integral over F: static}) for moving objects.
The EM field correlation function corresponding to the inside fluctuations is then obtained as
\begin{align}\label{Eq: <EE> in moving}
  \langle \bE (\omega,\bx) \otimes \bE^* (\omega,\by)\rangle_{\innf}=  &\frac{\omega^2}{4c^2}\sum_{\alpha_m}a_{\inn}(\omega-\Omega m) \left(1-|S_{\alpha_m}|^2\right) \bE^{\out}_{\alpha_m}(\omega,\bx)\otimes \bE^{\out*}_{\alpha_m}(\omega,\by).
\end{align}
 The total radiation per unit time can be obtained by integrating over the Poynting vector as
\begin{align}\label{Eq: em radiation}
  {\cal P}&= \frac{1}{4\pi}\int_{0}^{\infty} d\omega \, \omega \,\sum_{\alpha_m} \left(a_{\inn}(\omega-\Omega m)-a_{\out}(\omega)\right)\,  (1- |S_{\alpha_m}(\omega)|^2)\nonumber\\
  &=\int_{0}^{\infty} \frac{d\omega }{2\pi}\, \hbar \omega \,\sum_{\alpha_m} \left(n(\omega-\Omega m,T)-n(\omega,T_0)\right)\,  (1- |S_{\alpha_m}(\omega)|^2).
\end{align}
At zero temperature everywhere, $n(\omega-\Omega m,0)-n(\omega,0)=-\Theta(\Omega m -\omega)$, and the quantum radiation happens in the superradiating regime; see the discussion in Sec.~\ref{Sec: Radiation}. Again we note that our derivation leading to Eq.~(\ref{Eq: em radiation}) is not specific to a coordinate system and shape as long as the object is a solid of revolution with the angular momentum $m$ being a good quantum number. In a general basis where the scattering matrix is not diagonal (except in $m$), we have
\begin{align}\label{Eq: em trace formula}
  {\cal P}&= \int_{0}^{\infty} \frac{d\omega }{2\pi}\, \hbar \omega \,\sum_{\alpha_m,\beta_m} \left(n(\omega-\Omega m,T)-n(\omega,T_0)\right)\,  (\delta_{\beta_m \alpha_m}- |S_{\beta_m\alpha_m}(\omega)|^2)\nonumber \\
  &=\int_{0}^{\infty} \frac{d\omega }{2\pi}\, \hbar \omega \,\mbox{Tr}\left[ \left(n(\omega-\Omega \hat l_z,T)-n(\omega,T_0)\right)\,  (\mathbb I- \mathbb S^\dagger (\omega)\mathbb S(\omega))\right],
\end{align}
where $\hat l_z$ is the angular momentum operator. This equation casts the quantum (and thermal) radiation from a rotating object into a Trace formula applicable to any shape with rotational symmetry. For detailed discussions on thermal radiation and the heat transfer for arbitrary objects, see Ref.~\cite{kruger12}.

\subsection{A test object in the presence of a rotating body}\label{Sec: test object}
In this section, we study the interaction of the radiation field from a rotating body with a test object at rest, and assume that both objects are dielectric spheres. The overall EM field correlation function is given by the sum of Eqs.~(\ref{Eq: <EE> out moving}) and (\ref{Eq: <EE> in moving}) as
\begin{equation}\label{Eq: corr fn total}
    \langle \bE\otimes\bE^*\rangle=\langle \bE\otimes\bE^*\rangle_{\innf}+\langle \bE\otimes\bE^*\rangle_{\outf}.
\end{equation}
In the following, we consider the limit of zero temperature both in the object and the environment. The generalization to finite temperature is straightforward. Similar to Sec.~\ref{Sec: Scalar field}, the correlation function in Eq.~(\ref{Eq: corr fn total}) can be recast as
\begin{equation}\label{Eq: corr fn rad + non-rad}
    \langle \bE \otimes \bE^* \rangle =   \langle \bE \otimes \bE^* \rangle_{\rm non-rad}+ \langle \bE \otimes \bE^* \rangle_{\rm rad} \,,
\end{equation}
where we have broken up the correlation function into radiative (due to propagating photons) and non-radiative (due to zero-point fluctuations) parts. The latter is given by
\begin{align}
  \langle \bE (\omega, \bx)\otimes \bE^*(\omega,\by) \rangle_{\rm non-rad}= \hbar\,\sgn(\omega) \im \mathbb G(\omega,\bx,\by),
\end{align}
where $\mathbb G$ is the Green's function in the presence of a rotating object. This equation is reminiscent of the FDT in equilibrium; the  term in the RHS is purely real and thus does not contribute to the radiation, but leads to a Casimir-like force between the rotating body and nearby objects. The radiative term in the correlation function can be obtained from Eqs.~(\ref{Eq: corr fn total}) and (\ref{Eq: corr fn rad + non-rad}) as
\begin{align} \label{Eq: G-rad}
  \langle \bE (\omega, \bx)\otimes \bE^*(\omega,\by) \rangle_{\rm rad}\approx\frac{\hbar\omega^2}{2c^2} \sum_{\alpha_m}\Theta(\Omega m-\omega) \, \left(|S_{\alpha_m}|^2-1\right) \bE^{\out}_{\alpha_m}(\omega,\bx)\otimes \bE^{\out*}_{\alpha_m}(\omega,\by)\,,
\end{align}
and contributes to the Poynting vector in the superradiating regime $0<\omega<\Omega m$.

To find the interaction with a \emph{test }object, we only consider the radiative term in the correlation function for two reasons. First radiation pressure exerts a force falling off more slowly with the separation distance compared to the non-radiative part. Furthermore, non-radiative fluctuations give rise to a potential energy depending only on the separation distance akin to the Casimir energy. The test object being spherical, the corresponding tangential force or torque due to the corresponding term in Eq.~(\ref{Eq: corr fn rad + non-rad}) is identically zero.

The radiation from a (non-magnetic) rotating sphere is dominated by the lowest (electric) partial wave $(l=1,m=1,P=E)$ in which case Eq.~(\ref{Eq: G-rad}) yields
\begin{align}
  \langle \bE (\omega, \bx)\otimes \bE^*(\omega,\by) \rangle_{\rm rad}= \frac{\hbar\omega^2}{2c^2}  \,\Theta(\Omega-\omega)\left(|S_{11E}|^2-1\right) \bE^{\out}_{11E}(\omega,\bx)\otimes \bE^{\out*}_{11E}(\omega,\by).
\end{align}
The partial waves in Eq.~(\ref{Eq: G-rad}) are defined in spherical basis as
\begin{align}\label{Eq: spherical fns}
  &\bE^{\out}_{lmM}(\omega,\bx)=\frac{\sqrt{\omega/c}}{\sqrt{l(l+1)}} \nabla \times \, h_l^{(1)}\left(\frac{\omega r}{c}\right)\,Y_{lm}(\theta, \phi) \bx, \nonumber \\
  &\bE^{\out}_{lmE}(\omega,\bx)=-i\frac{\sqrt{c/\omega}}{\sqrt{l(l+1)}} \nabla \times \nabla \times \, h_l^{(1)}\left(\frac{\omega r}{c}\right)\,Y_{lm}(\theta, \phi) \bx,
\end{align}
where $Y_{lm}$ is usual spherical harmonic function, and $h_l^{(1)}$ is the spherical Hankel function of the first kind. The normalization is chosen to ensure the conditions in Eqs.~(\ref{Eq: identity}) and (\ref{Eq: identity 2}).

In order to find the scattering from the second object, we expand the EM field around its origin located at a separation $d$ {on} the $x$ axis. To the lowest order in frequency, we have
\begin{align}
  \bE^{\out}_{11E}(\omega,\bx)=\mathcal U_{11E, 11E}   \bE^{\reg}_{11E}(\omega,\tilde \bx) +\mathcal U_{10M, 11E}   \bE^{\reg}_{10M}(\omega,\tilde \bx)+\cdots,
\end{align}
where $\tilde \bx$ is defined with respect to the new origin. The {\it regular} functions are defined by replacing the spherical Hankel function $h^{(1)}_l$ in Eq.~(\ref{Eq: spherical fns}) by the spherical Bessel function $j_l$. The {\it translation }matrices are given by~\cite{Rahi09}
\begin{align}
   \mathcal U_{11E, 11E}= h^{(1)}_0\left(\frac{\omega d}{c}\right), \hskip .3in \mathcal U_{10M, 11E}= \frac{\sqrt{2}\omega d}{4c}h^{(1)}_0\left(\frac{\omega d}{c}\right).
\end{align}
Next we consider the scattering from the test object:
\begin{equation}
  \bE^{\reg}_{lmP}(\omega,\tilde \bx)\quad\rightarrow \quad \frac{1}{2}\left(\bE^{\inn}_{lmP}(\omega,\tilde\bx) +\mathfrak{S}_{lmP}\bE^{\out}_{lmP}(\omega,\tilde\bx)\right),
\end{equation}
where $\mathfrak{S}_{lmP}$ is the corresponding scattering matrix. We then find the EM field correlation function upon one scattering from the test object as
\begin{align}
  &\langle \bE (\omega, \tilde \bx)\otimes \bE^*(\omega,\tilde \by) \rangle_{\rm rad}=\frac{\hbar \omega^2}{8c^2}  \,\Theta(\Omega-\omega)\left(|S_{11E}|^2-1\right) \times\nonumber \\ &\left[\mathcal U_{11E, 11E}\left(\bE^{\inn}_{11E}(\omega,\tilde\bx) +\mathfrak{S}_{11E}\bE^{\out}_{11E}(\omega,\tilde\bx)\right)+\mathcal U_{10M, 11E}\left(\bE^{\inn}_{10M}(\omega,\tilde\bx) +\mathfrak{S}_{10M}\bE^{\out}_{10M}(\omega,\tilde\bx)\right)\right]\nonumber \\
  &\otimes \overline{\left[\mathcal U_{11E, 11E}\left(\bE^{\inn}_{11E}(\omega,\tilde\by) +\mathfrak{S}_{11E}\bE^{\out}_{11E}(\omega,\tilde\by)\right)+\mathcal U_{10M, 11E}\left(\bE^{\inn}_{10M}(\omega,\tilde\by) +\mathfrak{S}_{10M}\bE^{\out}_{10M}(\omega,\tilde\by)\right)\right]}\,.
\end{align}
Having the correlation functions, we can compute physical quantities of interest. In computing the torque, the partial waves $(1,1,E)$ and $(1,0,M)$ decouple; however, the latter does not contribute since its angular momentum along the $z$ axis is zero. We then find that the torque falls off as $1/d^2$ with the separation distance as
\begin{align}
  M&\sim \frac{\hbar }{8\pi} \int_{0}^{\Omega} {d\omega} \, (|S_{11E}|^2-1) \left|{\cal U}_{11E,11E}\right|^2(1-|\mathfrak{S}_{11E}|^2), \nonumber \\
   &=\frac{\hbar c^2}{8\pi d^2} \int_{0}^{\Omega} d\omega \, \frac{1}{\omega^2} (|S_{11E}|^2-1)(1-|\mathfrak{S}_{11E}|^2).
\end{align}
For small particles whose polarizability are $\alpha_1$ and $\alpha_2$ for rotating and static bodies, respectively, we find
\begin{align}
  M=\frac{8\hbar c^2}{9\pi d^2} \int_{0}^{\Omega} d\omega \, {\omega^4} |\im \alpha_1 (\omega-\Omega)|\, \im \alpha_2 (\omega).
\end{align}

Computing the force is more complicated since the two partial waves mix, and one has to find their overlap via the Maxwell stress tensor
\begin{equation}
  T_{ij}(\omega)=E_i(\omega) E_j^*(\omega)+B_i(\omega) B_j^*(\omega)-\frac{1}{2}\left(\bE^2+\bB^2\right)\delta_{ij}.
\end{equation}
The $y$-component of the force, perpendicular the $x$ axis connecting the two objects, and the $z$ axis along the rotation, is obtained as
\begin{equation}
  F_y=\int \frac{d\omega}{2\pi} \int r^2d\Omega_{\hat r} \,\langle T_{ij}\rangle\,\hat r_i \hat y_j,
\end{equation}
with $\Omega_{\hat r}$ being the solid angle corresponding to the unit vector $\hat r$ from the origin of the test object. A lengthy, though straightforward, calculation leads to
\begin{align}
      F_y&=    \frac{\hbar}{4\pi c^2}\int_{0}^{\Omega} {d\omega}\,{\omega^2} \,(|S_{11E}|^2-1) \,\, \mathcal U_{10M, 11E} \mathcal U_{11E, 11E} \,\mathcal T_{10M,11E} \,\re(-1+\overline{\mathfrak{S}_{10M}}\mathfrak{S}_{11E}),
\end{align}
where $\mathcal T_{10M,11E}$ characterizes the stress tensor sandwiched between the two partial waves, whose dependence on frequency is given by
\begin{equation}
  \mathcal T_{10M,11E} =-\frac{\pi c}{2\sqrt{2}\omega}.
\end{equation}
For a non-magnetic object, we can safely assume $\mathfrak{S}_{10M}\approx1$ since its frequency dependence can be neglected compared to $\mathfrak{S}_{11E}$, hence
\begin{align}
F_y=\frac{\hbar}{32\pi d}\int_{0}^{\Omega} d\omega \,(|S_{11E}|^2-1) (1-\re\mathfrak{S}_{11E}).
\end{align}
Notice that the force falls off as the inverse separation distance while the usual Casimir force decays much faster.
For small particles with polarizability $\alpha_1$ and $\alpha_2$, we find
\begin{align}
  F_y=\frac{\hbar }{9\pi d} \int_{0}^{\Omega} d\omega \, {\omega^6} |\im \alpha_1 (\omega-\Omega)|\, \im \alpha_2 (\omega).
\end{align}

\subsection{Vacuum friction on a rotating object}\label{Sec: Friction on rotating objects}
For a rotating object, we have to solve a complicated equation, Eq.~(\ref{rel em inside sln}), but in the lowest order in velocity we can neglect the explicit dependence on velocity and set $\mathbb D\approx \tilde {\mathbb D}\approx \mathbb I$, changing the argument of the dielectric function as $\epsilon(\omega) \to \epsilon(\omega-\Omega m)$.
Finally, at zero temperature, only frequencies within the range $[0,\Omega]$ contribute. We examine the cases where the object is a sphere or a right circular cylinder.

\subsubsection{Sphere}
EM Scattering from a sphere is most conveniently described in a basis $(l,m, P)$ where $l$ corresponds to the total angular momentum, $m$ is the angular momentum along the $z$ axis, and $P$ is the polarization. In the approximations made here, the lowest partial wave, $l=1$, gives the leading order, while larger $l$s are suppressed by higher powers of the (linear) velocity divided by the speed of light. We assume a non-magnetic object, thus the electric polarizability gives the leading contribution to scattering matrix as
\begin{equation}
  S_{1mE}(\omega)=1+i \frac{4\omega^3 }{3c^3}\, \alpha(\omega-\Omega m),
\end{equation}
where $\alpha(\omega)$ is the polarizability of a spherical object at low frequencies (appropriate to the problem of a rotating object at a small angular velocity) depending solely on the dielectric function $\epsilon(\omega)$ which is assumed to be a homogenous but frequency-dependent function within the object. Note that, at zero temperature, only $m=1$ (and not $m=0,-1$) contributes to the radiation.

\emph{Radiation}: The rate of energy radiation to the vacuum is obtained as
\begin{align}
  {\cal P}&\approx\int_{0}^{\Omega} \frac{d\omega}{2\pi}\, \hbar \omega\left(|S_{11E}|^2-1\right) \nonumber \\
   &\approx\frac{4\hbar}{3\pi c^3}\int_{0}^{\Omega} d\omega \,\omega^4 \, \left(-\im \alpha(\omega-\Omega)\right),
\end{align}
where we have kept only the leading term in powers of frequency.
For a dielectric sphere of radius $R$, the polarizability is $
\alpha(\omega)=R^3{(\epsilon(\omega)-1)}/{(\epsilon(\omega)+2)}$; the radiation is then given by~\cite{Manjavacas10}
\begin{align}
  {\cal P}\approx\frac{4\hbar R^3}{3\pi c^3}\int_{0}^{\Omega} d\omega \,\omega^4 \, \left|\im \frac{\epsilon(\omega-\Omega)-1}{\epsilon(\omega-\Omega)+2}\right|.
\end{align}
Similarly, the frictional torque is obtained as
\begin{align}
  M\approx\frac{4\hbar R^3}{3\pi c^3}\int_{0}^{\Omega} d\omega \,\omega^3 \, \left|\im \frac{\epsilon(\omega-\Omega)-1}{\epsilon(\omega-\Omega)+2}\right|.
\end{align}
For a metallic particle, $\epsilon=1+i4\pi \sigma/\omega$, and $\im \alpha\approx 3\omega R^3/4\pi \sigma$. Hence, we find~\cite{Manjavacas10}
\begin{align}
  {\cal P}= \frac{\hbar R^3 \Omega^6}{30\pi^2 c^3 \sigma}\,, \qquad
  {M}= \frac{\hbar R^3 \Omega^5}{20\pi^2 c^3 \sigma}.
\end{align}

\emph{Entropy generation}: Making the above approximation, the rate of entropy generation is given by
\begin{align}
  {\cal S}&\approx -k_B \int_{0}^{\Omega} \frac{d\omega}{2\pi}\, \left(|S_{11E}|^2-1\right)\log \left(|S_{11E}|^2-1\right) \nonumber \\
  &=k_B \, {\cal N}_{11E} \,|\log \frac{ R^3 \Omega^4}{ c^3 \sigma}|, \qquad \mbox{with} \qquad {\cal N}_{11E}=\frac{R^3 \Omega^5}{30\pi^2 c^3 \sigma},
\end{align}
where the second line is computed for a metallic particle, and ${\cal N}_{11E}$ is the total of number of photons with the quantum number $11E$ radiated per unit time.

\emph{Uncertainty in angular momentum}---For small angular velocities, one can compute the uncertainty of angular momentum from Eq.~(\ref{Eq: uncertainty}). Notice that $M \propto \Omega^5$, thus
\begin{equation}\label{Eq: uncertainty 2}
  I \Delta \Omega \approx \sqrt{\frac{\hbar I \Omega_0}{5}}.
\end{equation}

\subsubsection{Cylinder}
For a cylinder, the scattering matrices are more complicated due to mixing between the two polarizations. A complete basis for cylindrical waves is $( m,k_z, P)$ with $k_z$ being the wavevector parallel to the $z$ axis, and polarizations labelled by $P$. In the limit of a thin or slowly rotating cylinder where  $\Omega R/c, \epsilon \, \Omega R/c \ll 1$, the first partial wave, $m=1$, gives the leading contribution while $k_z$ should be integrated over all propagating waves. The corresponding scattering matrices are
\begin{align}
  S_{1k_zMM}(\omega)&= 1+\frac{i\pi}{2} \frac{\epsilon(\omega-\Omega)-1}{\epsilon(\omega-\Omega)+1} \, \frac{\omega^2 }{c^2} R^2,\nonumber\\
  S_{1k_zEE}(\omega)&= 1+\frac{i\pi}{2} \frac{\epsilon(\omega-\Omega)-1}{\epsilon(\omega-\Omega)+1} \, k_z^2 R^2, \nonumber\\
  S_{1k_zEM}(\omega)=S_{1k_zME}(\omega)&= \frac{i\pi}{2} \frac{\epsilon(\omega-\Omega)-1}{\epsilon(\omega-\Omega)+1} \frac{\omega k_z }{c} R^2,
\end{align}
where the argument of the dielectric function is $\omega-\Omega$ corresponding to $m=1$. Again we have assumed that the object is described by a spatially constant but frequency-dependent $\epsilon(\omega)$.

\emph{Radiation}:
The energy radiation per unit time is obtained as~\cite{Maghrebi12}
\begin{align}\label{Eq: rad from cylinder}
  {\cal P}
  &  \approx\int_{0}^{\Omega} \frac{d\omega}{2\pi}\, \hbar \omega \int_{-\omega/c}^{\omega/c} \frac{L dk_z}{2\pi}\sum_{P,P'\in\{M, E\}} \left[ \left|S_{1k_zPP'}(\omega)\right|^2-\delta_{PP'} \right] \nonumber \\
  &\approx\frac {2 \hbar L R^2}{3\pi c^3}\int_{0}^{\Omega} d\omega \, \omega^4 \left|\im \frac{\epsilon(\omega-\Omega)-1}{\epsilon(\omega-\Omega)+1}\right|,
\end{align}
where we have neglected terms of the order of $R^4$. If the cylinder has a small conductivity described by the dielectric function $\epsilon=1+i4\pi\sigma/\omega$ with $\sigma \ll \Omega$, Eq.~(\ref{Eq: rad from cylinder}) yields
\begin{equation}
  {\cal P}=\frac{8\hbar LR^2 \Omega^4 \sigma}{c^3} \log\frac{\Omega}{\sigma},
\end{equation}
in agreement with the results of Ref.~\cite{Zel'dovich86}.

In the opposite limit where $\Omega \ll \sigma$, we find
\begin{align}
  {\cal P}= \frac{\hbar L R^2 \Omega^6}{90\pi^2 c^3 \sigma}\,, \qquad
  {M}= \frac{\hbar L R^2 \Omega^5}{60\pi^2 c^3 \sigma}.
\end{align}
The radiation from a rotating cylinder indeed takes a similar form to that of a rotating sphere; however, while $\Omega R/c$ is constrained to be small in the nonrelativistic limit, $\Omega L/c$ is not.

The results for entropy generation and angular-momentum uncertainty also bear a close resemblance to a rotating sphere. Specifically, with $M \sim \Omega^5$, we find the same relation in Eq.~(\ref{Eq: uncertainty 2}) for a rotating cylinder. One may speculate that this result holds for any rotating object with an arbitrary geometry as long as the angular velocity is small.

To get an estimate for {the magnitude of} radiation effects, {we} consider a {rapidly} spinning nanotube of radius $R$ and length $L$, and assume that $\Omega R/c$ is small. We then find that the rotation slows down by an order of magnitude over a time scale of $\tau\sim (I/\hbar ) \, {(}c^3 /L R^2\Omega^3{)}$.
  The moment of inertia of a nanotube can be as small as $10^{-33}$ in SI units~\cite{Sadeghpour02} (compare with $\hbar \approx 10^{-34}$). So even at small velocities, $\tau$ can be of the order of a few hours.

\begin{table}[h]
\begin{ruledtabular}
\begin{tabular}{c|cc}
&Scalar& EM\\
\hline\hline
Free Lag.& $\frac{\omega^2}{c^2}| \Phi_\omega| ^2 -|\nabla\Phi_\omega|^2$ & $|\bE_\omega|^2-\frac{c^2}{\omega^2}|\nabla \times \bE_\omega|^2$ \\
\hline
Current density & $\im \Phi_\omega^* \nabla \Phi_\omega$ & $ \frac{c^2}{\omega^2}\im \bE_\omega^* \times (\nabla \times \bE_\omega) $ \\
\hline
Medium's Lag. & $(\epsilon (\omega)-1)\frac{\omega^2}{c^2}| \Phi_\omega|^2$ & $(\epsilon(\omega)-1)|\bE_\omega|^2$ \\
\hline
Random source Lag. & $- i\frac{\omega}{c} \Phi^*_\omega \varrho_\omega $ & $\bE_\omega^*\cdot\bK_\omega$ \\
\hline
Fluctuation-Dissipation & $\langle \varrho_\omega(\bx) \varrho^*_\omega(\by) \rangle =a(\omega)\im \epsilon(\omega) \, \delta(\bx-\by)$ & $\langle \bK_\omega(\bx) \otimes \bK_\omega^*(\by)\rangle =a(\omega) \im \epsilon(\omega) \,\delta(\bx-\by) \mathbb I$ 
\end{tabular}
\end{ruledtabular}
\caption{\label{table:1} Comparison of scalar and electromagnetic formulations for a static medium---The Lagrangian in free space, the current density in vacuum, the Lagrangian terms due to the linear response of the medium and fluctuating sources, and the fluctuation-dissipation relation for scalar and electromagnetic fields respectively. $a(\omega)=\hbar \, \coth\left({\hbar \omega}/{2k_B T}\right)$. }
\end{table}

\begin{table}[h]
\begin{ruledtabular}
\begin{tabular}{c|cc}
&Scalar& EM\\
\hline\hline
Medium's Lag. & $(\epsilon (\omega')-1)\frac{\omega'^2}{c^2}| \Phi_\omega|^2$ & $(\epsilon(\omega')-1)|\bE_\omega|^2$ \\
\hline
Random source Lag. & $- i\frac{\omega'}{c} \Phi^*_\omega \varrho_\omega $ & $(\bE_\omega^*+\frac{i}{\omega}\bv\times (\nabla\times \bE_\omega^*))\cdot\bK_\omega$ \\
\hline
Fluctuation-Dissipation & $\langle \varrho_\omega(\bx) \varrho^*_\omega(\by) \rangle =a(\omega')\im \epsilon(\omega') \, \delta(\bx-\by)$ & $\langle \bK_\omega(\bx) \otimes \bK_\omega^*(\by)\rangle =a(\omega') \im \epsilon(\omega') \,\delta(\bx-\by) \mathbb I$ 
\end{tabular}
\end{ruledtabular}
\caption{\label{table:2} Comparison of scalar and electromagnetic formulations for a rotating medium---The last three rows of Table~\ref{table:1} (pertaining to the interior of the medium) are modified under steady rotation. Note that $\omega'=\omega-\Omega m$ is the shifted frequency in the comoving frame rotating at the rate $\Omega$.}
\end{table}

\section*{Acknowledgements}
This work was supported by the U.\ S.\ Department of Energy under cooperative research agreement \#DF-FC02-94ER40818 (MFM and RLJ), NSF Grant No. DMR-12-06323 (MK).

\section*{Appendix A: Green's theorem}\label{appen}

The vector Green's theorem reads
\begin{equation}
  E_i(\bx)=\oint d\mathbf\Sigma\cdot \left[ (\nabla \times \bG_i(\bx,\bz))\times \bE (\bz) +\bG_i(\bx,\bz)\times (\nabla \times\bE(\bz)) \right],
\end{equation}
where $E_i$ is the $i$ component of the electric field $\bE$ which satisfies the vector Helmholtz equation, and $\bG_i$ is a vector defined from the dyadic Green's function as $(\bG_i)_j=\mathbb G_{i j}$. Also note that the point $\bx$ is enclosed by the boundary of the integration. We choose $\bE=\bE^{\reg}_\beta$, a partial wave indexed by $\beta$, and also employ the definition of the Green's function in Eq.~(\ref{Eq: G out-out0}) to find
\begin{equation}
  [\bE^{\reg}_\beta(\bx)]_i=i\sum_{\alpha} (\bE^{\reg}_\alpha(\bx))_i \oint d\mathbf\Sigma\cdot \left[ (\nabla \times \bE^{\out}_{\bar \alpha}(\bz))\times \bE^{\reg}_\beta(\bz) +\bE^{\out}_{\bar\alpha}(\bz)\times (\nabla \times\bE^{\reg}_{\beta}(\bz)) \right].
\end{equation}
The vector fields $\bE^{\reg}_{\alpha}$ constitute a complete set, hence
\begin{equation}
  i \oint d\mathbf\Sigma\cdot \left[ (\nabla \times \bE^{\out}_{\bar \alpha}(\bz))\times \bE^{\reg}_\beta(\bz) +\bE^{\out}_{\bar\alpha}(\bz)\times (\nabla \times\bE^{\reg}_{\beta}(\bz)) \right]=\delta_{\alpha \beta}.
\end{equation}
Using the definition of the regular wave-functions, we arrive at Eq.~(\ref{Eq: identity}).

\end{document}